\documentclass[fleqn,usenatbib,useAMS]{rasti}
\usepackage{hyperref}
\usepackage{times, amsmath, amssymb, amsfonts, latexsym}
\usepackage{fleqn, multirow, graphicx, xcolor, enumitem}
\usepackage{array, mathtools,subfigure, esint, multicol,caption}
\usepackage{soul,bm, pdflscape, algorithm, algpseudocode}

\DeclareCaptionLabelFormat{noname}{#2}
\newcommand{\Lim}[1]{\raisebox{0.5ex}{\scalebox{0.8}{$\displaystyle \lim_{#1}\;$}}}
\makeatletter
\@for\next:={int,iint,iiint,iiiint,dotsint,oint,oiint,sqint,sqiint,
	ointctrclockwise,ointclockwise,varointclockwise,varointctrclockwise,
	fint,varoiint,landupint,landdownint}\do{%
	\expandafter\edef\csname\next\endcsname{%
		\noexpand\DOTSI
		\expandafter\noexpand\csname\next op\endcsname
		\noexpand\ilimits@
	}%
}
\makeatother

\graphicspath{{./}{figures/}}

\title[Fast 1D CWT algorithms]{
	Comparisons between fast algorithms for the continuous wavelet transform and applications in cosmology: the 1D case
}
\author[Wang \& He ]{Yun Wang$^1$ and Ping He$^{1,2}$\thanks{E-mail: hep@jlu.edu.cn} \\
	$^{1}$College of Physics, Jilin University, Changchun 130012, P.R. China. \\
	$^{2}$Center for High Energy Physics, Peking University, Beijing 100871, P.R. China.}

\date{Accepted 06/06/2023. Received 15/05/2023; in original form 09/02/2023}

\begin{document}

\maketitle

\begin{abstract}
The continuous wavelet transform (CWT) is very useful for processing signals with intricate and irregular structures in astrophysics and cosmology. It is crucial to propose precise and fast algorithms for the CWT. In this work, we review and compare four different fast CWT algorithms for the 1D signals, including the FFTCWT, the V97CWT, the M02CWT, and the A19CWT. The FFTCWT algorithm implements the CWT using the Fast Fourier Transform (FFT) with a computational complexity of $\mathcal{O}(N\log_2N)$ per scale. The rest algorithms achieve the complexity of $\mathcal{O}(N)$ per scale by simplifying the CWT into some smaller convolutions. We illustrate explicitly how to set the parameters as well as the boundary conditions for them. To examine the actual performance of these algorithms, we use them to perform the CWT of signals with different wavelets. From the aspect of accuracy, we find that the FFTCWT is the most accurate algorithm, though its accuracy degrades a lot when processing the non-periodic signal with zero boundaries. The accuracy of $\mathcal{O}(N)$ algorithms is robust to signals with different boundaries, and the M02CWT is more accurate than the V97CWT and A19CWT. From the aspect of speed, the $\mathcal{O}(N)$ algorithms do not show an overall speed superiority over the FFTCWT at sampling numbers of $N\lesssim10^6$, which is due to their large leading constants. Only the speed of the V97CWT with real wavelets is comparable to that of the FFTCWT. However, both the FFTCWT and V97CWT are substantially less efficient in processing the non-periodic signal because of zero padding. Finally, we conduct wavelet analysis of the 1D density fields, which demonstrate the convenience and power of techniques based on the CWT. We publicly release our CWT codes as resources for the community.
\end{abstract}


\section{Introduction}
\label{sec:intro}

Wavelets are wave-like functions that are localized in both the real and Fourier domains. Hence, by convolving a signal under investigation with the dilated or contracted wavelets, the local features at various scales will be extracted, and this process is called wavelet transform \citep[WT; e.g.,][]{Daubechies1992, Kaiser1994, Addison2017}. There are two basic types of WT: discrete WT (DWT) and continuous WT (CWT). The DWT, using orthogonal wavelets, operates over coarse dyadic scales and positions. In contrast to the DWT, the CWT offers a highly redundant representation of the signal, which ensures that intricate structures or textures can be resolved quite well \citep{Addison2018}. However, the redundancy also makes the direct computation of the CWT terribly inefficient, which requires a time complexity of $\mathcal{O}(N^2)$ per scale, where $N$ is the number of data points. One more efficient way to implement CWT is to use the Fast Fourier Transform (FFT) with a complexity $\mathcal{O}(N\log_2N)$, since the convolution in the real domain is equivalent to multiplication in the frequency domain, i.e. convolution theorem \citep{Torrence1998, Antonio2004, Press2007, Arts2022}.

Consequently, the CWT is becoming increasingly popular in many areas of science and engineering. In the context of astrophysics and cosmology, the CWT has been used for various studies including but not limited to, identifying structures and substructures from the galaxy catalogue \citep[e.g.,][]{Slezak1990, Slezak1993, Escalera1992, Escalera1994, Flin2006, Schwinn2018}, analyzing the fractal properties of the galaxy distribution \citep[e.g.,][]{Martinez1993, Rozgacheva2012}, analyzing galactic images \citep[e.g.,][]{Frick2001, Frick2016, Tabatabaei2013, Robitaille2014, Arshakian2016}, detecting baryon acoustic oscillation features \citep[e.g.,][]{Tian2011, Arnalte-Mur2012, Labatie2012}, investigating the turbulence in the intracluster medium \citep[e.g.,][]{Shi2018, Roh2019} and characterizing the cosmic density fields at low redshifts \citep[e.g.,][]{Wang2022a, Wang2022b}.

Since the COBE detection of the CMB anisotropy in 1992, cosmology has emerged as a precision, data-driven science \citep{Turner2022}. The observational experiments such as the Euclid space mission \citep[Euclid,][]{Laureijs2011}, the Dark Energy Spectroscopic Instrument \citep[DESI,][]{Levi2013}, and the Square Kilometre Array \citep[SKA,][]{Bacon2020}, and state-of-the art cosmological simulations such as the IllustrisTNG \citep{Pillepich2018}, the SIMBA \citep{Dave2019}, and the MillenniumTNG \citep{Hernandez-Aguayo2022}, are producing increasingly growing amounts of data, which need to be analyzed by high-performance algorithms and methods. Therefore, the fast CWT algorithms with $\mathcal{O}(N)$ complexity are obviously more attractive than the FFT-based implementation of the CWT (FFTCWT) with $\mathcal{O}(N\log_2N)$ complexity. Fortunately, a great effort has been made to develop fast CWT algorithms without using the FFT \citep[e.g.,][]{Unser1994, Berkner1997, Vrhel1997, Munoz2002, Omachi2007, Arizumi2019}, which achieve the time complexity of $\mathcal{O}(N)$ per scale. However, some $\mathcal{O}(N)$ algorithms are only applicable to particular cases. For example, the algorithm of \citet{Unser1994} is restricted to integer scales, the algorithm of \citet{Berkner1997} is only available for wavelets which are derivatives of the Gaussian function, and the algorithm of \citet{Omachi2007} is only applicable for polynomial wavelets.

What we need are fast CWT algorithms with no restrictions on the wavelet, and with arbitrarily fine scale resolution. Therefore, in this study, we will consider the $\mathcal{O}(N)$ algorithms proposed by \citet{Vrhel1997}, by \citet{Munoz2002} and by \citet{Arizumi2019}. For convenience, we denote these three algorithms as the V97CWT, the M02CWT, and the A19CWT, respectively. The V97CWT is a fast recursive algorithm based on the finite impulse response (FIR) and infinite impulse response (IIR) filtering techniques with filter coefficients determined by two compactly supported auxiliary functions. The M02CWT reaches the linear complexity by decomposing both the wavelet and the signal into B-splines, and the A19CWT approximates the wavelet as piecewise polynomials and reduces the number of operations using integration by parts.

Motivated by the facts that (1) all these powerful algorithms are 1D, and (2) there is no any publicly available source code for the V97CWT, M02CWT, and A19CWT algorithms, we must conduct a systematic comparison study of them to benchmark their actual performance, which is the basis for developing high-dimensional fast CWT algorithms to analyze high-dimensional data, e.g. the 2D weak-lensing maps and 3D spatial distribution of matter. For some simple 1D functions, such as sine, cosine, and Gaussian functions, their CWTs can be evaluated by analytical calculations. So the accuracy of their numerical CWTs can be verified by the corresponding analytical results. Finally, it should be noted that the CWT for 1D signals is not trivial in astrophysics and cosmology, as it is also applicable to a wide range of scenarios, such as analyzing the light curves of astronomical sources \citep[e.g.,][]{Tarnopolski2020, Ren2022}, subtracting the foreground emission from the 21 cm signal \citep[e.g.,][]{Gu2013, Li2019}, measuring the small-scale structure in the Lyman-$\alpha$ forest \citep[e.g.,][]{Lidz2010, Garzilli2012, Wolfson2021}, investigating the time-frequency properties of the gravitational waves \citep[e.g.,][]{Tary2018}, characterizing the 1D density fields \citep[e.g.,][]{daCunha2018, Wang2021, Wang2022a}, and so on. We publicly release the Fortran 95 implementations\footnote{The Fortran 95 codes are available at \url{https://github.com/WangYun1995/FortranCWT}} and their Python wrappers\footnote{The Python wrappers are available at \url{https://github.com/WangYun1995/pyFortranCWT}} of the fast CWT algorithms described in this manuscript, in the hope that the community will use them to perform wavelet analysis of 1D signals.

The paper is organized as follows. We briefly introduce the mathematical formalism of the CWT in Section \ref{sec:cwt_form}. We review the fast CWT algorithms in Section \ref{sec:fast_cwt_algos}, and compare the performance between them in Section \ref{sec:performance}. We present simple applications of the 1D CWT in cosmology in Section \ref{sec:cosmo_apps}. Finally, in Section \ref{sec:conclusions}, we summarize our main findings and present the conclusions.

For convenience of the readers, in Table \ref{tab:notations}, we list the acronyms frequently used in our paper, with their meanings explained.

\begin{table}
	\centering
	\caption{The acronyms frequently used in the paper, with their meanings explained.}
	\label{tab:notations}
	\begin{tabular}{ll}
		\hline
		Acronym & Meaning  \\ \\[-1.1em]
		\hline
		CWT      & continuous wavelet transform                    \\
		ICWT     & inverse continuous wavelet transform            \\
		CBSW     & cubic B-spline wavelet                          \\
		GDW      & Gaussian-derived wavelet                        \\
		CW-GDW   & cosine-weighted Gaussian-derived wavelet        \\
		MW       & Morlet wavelet                                  \\
		FT       & Fourier transform                               \\
        FFT      & Fast Fourier transform                          \\
        FFTCWT   & the fast CWT algorithm based on the FFT         \\
        V97CWT   & the fast CWT algorithm of \citet{Vrhel1997}     \\
        M02CWT   & the fast CWT algorithm of \citet{Munoz2002}     \\
        A19CWT   & the fast CWT algorithm of \citet{Arizumi2019}   \\
   		\hline
	\end{tabular}
\end{table}

\section{The Formalism of the Continuous Wavelet Transform}
\label{sec:cwt_form}

\begin{figure*}
	\centerline{\includegraphics[width=0.95\textwidth]{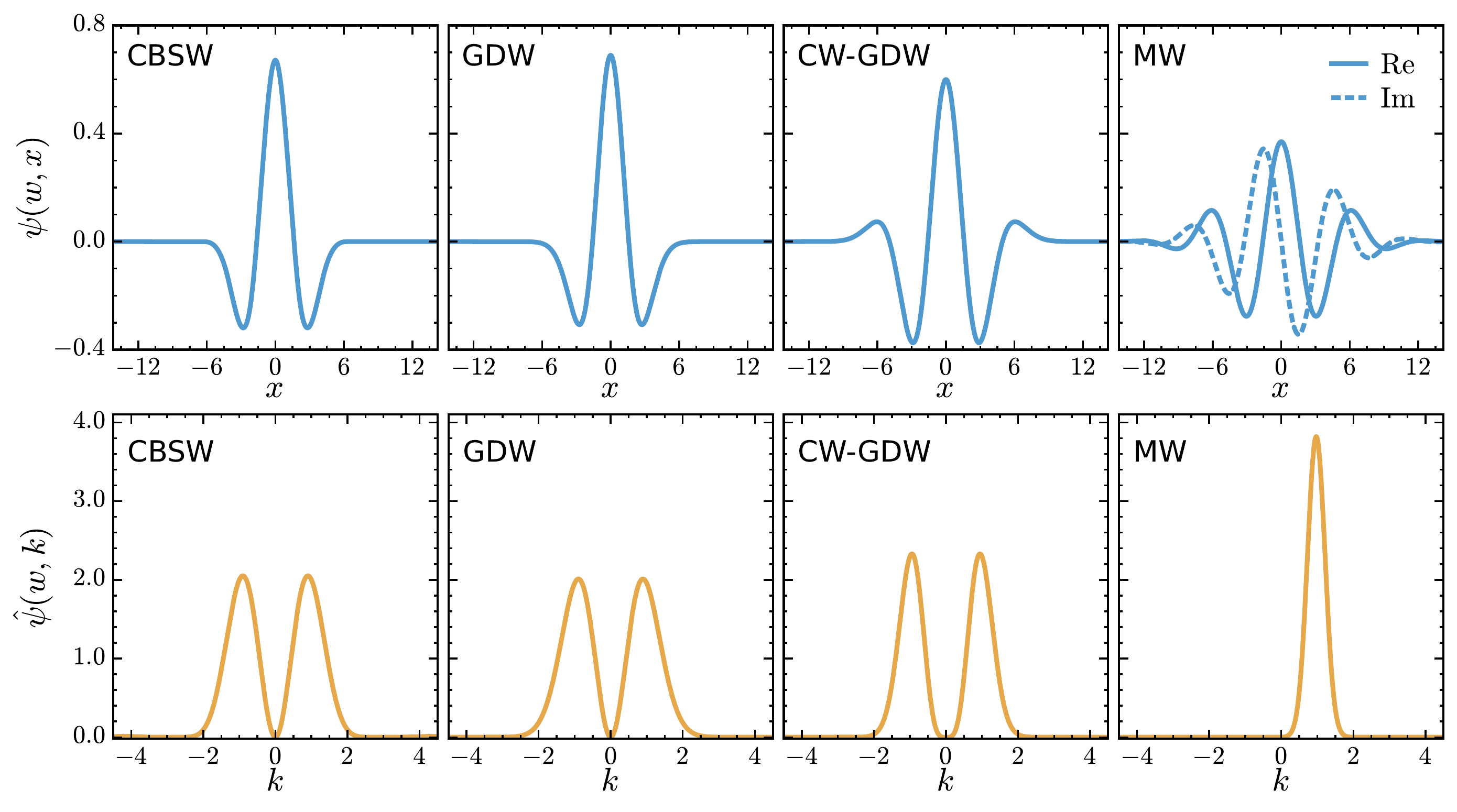}}
	\caption{\textit{Top row}: the CBSW, GDW, CW-GDW and MW in the real domain, at scale $w=c_w$, i.e. $k_\mathrm{pseu}=1$. \textit{Bottom row}: the respective Fourier transforms of the wavelets in the top row.  }
	\label{fig:wavelets}
\end{figure*}

The CWT $W_f(w,x)$ of a 1D real signal $f(x)$ is defined as the convolution of $f(x)$ with a scaled wavelet, i.e.
\begin{equation}
\label{eq:CWT}
W_f(w,x)=\int_{-\infty}^{+\infty}f(u)\psi(w,x-u)\mathrm{d}u,
\end{equation}
where $w$ is the scale parameter with dimension of $[x]^{-1}$, and
\begin{equation}
\label{eq:scaled_wavelet}
\psi(w,x)=\sqrt{w}\psi(wx)
\end{equation}
is the scaled version of the mother wavelet
\begin{equation}
\label{eq:mother_wavelet}
\psi(x)=\psi(1,x).
\end{equation}
There are many different choices for the mother wavelet. In this study, we consider four kinds of wavelets: the cubic B-spline wavelet \citep[CBSW,][]{Munoz2002}, the Gaussian-derived wavelet \citep[GDW,][]{Wang2021}, the cosine-weighted Gaussian-derived wavelet \citep[CW-GDW,][]{Wang2022b}, and the Morlet wavelet \citep[MW,][]{Addison2017}. Table \ref{tab:wavelets} shows their formulas and properties, and Fig. \ref{fig:wavelets} gives a graphical representation.

As well known, the classical inverse CWT (ICWT) formula is a double integral over scale and space \citep[see e.g.][]{Addison2017}. In fact, there are simpler inverse ways. If the complex wavelet satisfies $\hat\psi(k)=0$ for $k<0$ and $0<|\mathcal{K}_\psi|<\infty$, where $\mathcal{K}_\psi=\int_0^{+\infty}\frac{\hat{\psi}(k)}{k}\mathrm{d}k$, then the original signal can be reconstructed by the known Morlet formula \citep[see e.g.][]{Shensa1993,Daubechies2011} as follows
\begin{equation}
\label{eq:complex_ICWT}
f(x)=\bar{f}+2\mathrm{Re}\left\{\frac{1}{\mathcal{K}_\psi}\int_0^{+\infty}\frac{W_f(w,x)}{\sqrt{w}}\mathrm{d}w\right\},
\end{equation}
where $\mathrm{Re}\{\ldots\}$ denotes the real part, and $\bar{f}= \Lim{L\rightarrow\infty}\frac{1}{L}\int_{-L/2}^{L/2}f(x)\mathrm{d}x$ is the average value of $f(x)$ over all space. If the real wavelet satisfies $\psi(x)=\psi(-x)$ and $0<|\mathcal{K}_\psi|<\infty$, then a single integral ICWT formula also exists, which is
\begin{equation}
\label{eq:real_ICWT}
f(x)=\bar{f}+\frac{1}{\mathcal{K}_\psi}\int_0^{+\infty}\frac{W_f(w,x)}{\sqrt{w}}\mathrm{d}w,
\end{equation}
Note that Equation \eqref{eq:real_ICWT} is the generalization of the inverse formula in \citet{Wang2021} and \citet{Wang2022a}, which holds for wavelets derived from the smoothing window function. We refer to Appendix \ref{sec:ICWT} for the derivation of Equation \eqref{eq:real_ICWT}.

\begin{table*}
	\centering
	\caption{Four mother wavelet functions and their properties. $\hat\psi(k)$ is the FT of $\psi(x)$, $C_\mathrm{N}$ is the normalization constant that makes $\int_{-\infty}^{+\infty}|\psi(x)|^2\mathrm{d}x=1$, $\mathcal{K}_\psi=\int_0^{+\infty}\frac{\hat\psi(k)}{k}\mathrm{d}k$ is a constant that ensures the existence of the single integral ICWT formula, $\chi$ is the half width of the wavelet's support $[-\chi, \chi]$, and $c_w=w/k_\mathrm{pseu}$ is the ratio between the wavelet scale and the corresponding pseudo Fourier frequency (see \citet{Wang2022a} for the definition of $c_w$). Note the GDW, CW-GDW and MW are not compactly supported, but decay exponentially. For these three wavelets, we set the values of $\chi$ to confirm $\int_{-\infty}^{+\infty}|\psi(x)|^2\mathrm{d}x\approx1$ and $\psi(x)\approx 10^{-14}$.}
	\label{tab:wavelets}
	\noindent
	\resizebox{\textwidth}{!}{
		\begin{tabular}{@{}ccccccc}
			\hline
			   & $\psi(x)$ & $\hat\psi(k)$ & $C_\mathrm{N}$ & $\mathcal{K}_\psi$ & $\chi$ &  $c_w$\\ \\[-1.2em]
			\hline
			CBSW    & $C_\mathrm{N}\big(2\beta^3(x)-\beta^3(x+1)-\beta^3(x-1) \big)\ ^a$ & $64C_\mathrm{N}\sin^6(k/2)/k^4$ & $\sqrt{30/31}$ & $\frac{1}{2}\sqrt{\frac{15}{62}}(27\ln3-32\ln2)$ & $3$ & $0.46609\ ^b$ \\ \\[-0.5em]
			GDW    & $C_\mathrm{N}(2-x^2)e^{-\frac{x^2}{4}}$ & $8\sqrt{\pi}C_\mathrm{N}k^2e^{-k^2}$ & $1/(18\pi)^{1/4}$ & $2\left(8\pi/9\right)^{1/4}$ & $12$ & $2/\sqrt{5}$\\ \\[-0.5em]
			CW-GDW  & $C_\mathrm{N}\big( (1-x^2)\cos x-x\sin x\big)e^{\frac{1-x^2}{2}}$ & $\sqrt{2\pi}C_\mathrm{N}k(k\cosh k-\sinh k)e^{-\frac{k^2}{2}}$ &  $\sqrt{\frac{8}{1+5e}}/\pi^{1/4}$ & $4\pi^{1/4}/\sqrt{1+5e}$ & $8$ & $0.42822\ ^c$ \\ \\[-0.5em]
			MW
			  & $C_\mathrm{N}(e^{-4\mathrm{i}x}-e^{-8})e^{-\frac{x^2}{2}}$ & $\sqrt{2\pi}C_\mathrm{N}e^{-8}(e^{4k}-1)e^{-\frac{k^2}{2}}$ & $\frac{e^8}{\sqrt{1-2e^4+e^{16}}}/\pi^{1/4}$ & $1.27484\ ^d$ & $7.5$ & $0.24264\ ^e$ \\ \\[-0.5em]
			\hline
			\multicolumn{4}{l}{$^a$ see Appendix \ref{sec:bsplines} for the definition of B-splines.}\\
			\multicolumn{4}{l}{$^b$ $0.466094761079290$}\\
			\multicolumn{4}{l}{$^c$ $0.428218886729052$ }\\
			\multicolumn{4}{l}{$^d$ $1.274837568937901$}\\
			\multicolumn{4}{l}{$^e$ $0.242640671273266$}\\
		\end{tabular}
	}
\end{table*}

\section{Fast Algorithms for the 1D CWT}
\label{sec:fast_cwt_algos}

For a discrete signal $f(n)\equiv f(n\Delta x)$ with sampling interval $\Delta x$, the CWT will be discretized in the following form:
\begin{equation}
\label{eq:discreteCWT}
W_f(\tilde{w},n)=\sqrt{\tilde{w}\Delta x}\sum_mf(m)\psi[\tilde{w}(n-m)],
\end{equation}
where $\tilde{w}=w\Delta x$ is the dimensionless scale parameter. It is clear that the computation of Equation \eqref{eq:discreteCWT} requires $N^2$ multiplications and additions per scale, where $N$ is the number of sampling points. Therefore, the high computational complexity makes this algorithm impractical for use. Next we will review four CWT algorithms with better performance, namely the FFTCWT, the V97CWT \citep{Vrhel1997}, the M02CWT \citep{Munoz2002}, and the A19CWT \citep{Arizumi2019}.
\subsection{FFTCWT}
\label{sec:fftcwt_algo}

If the discrete signal $f(n)$ is periodic with period $L=N\Delta x$, then it can be decomposed into a Fourier series as follows
\begin{equation}
\label{eq:Fourier_series_signal}
f(n)=\frac{1}{L}\sum_m\hat f(m)e^{-2\pi\mathrm{i} mn/N},
\end{equation}
where the Fourier transform (FT) $\hat f(m)$ is defined as
\begin{equation}
\label{eq:FT_signal}
\hat f(m)=\frac{L}{N}\sum_n f(n)e^{2\pi\mathrm{i} mn/N}.
\end{equation}

By substituting Equation \eqref{eq:Fourier_series_signal} into Equation \eqref{eq:discreteCWT}, we get
\begin{equation}
\label{eq:discrete_CWT_conv}
W_f(\tilde{w},n)=\frac{1}{L}\sum_m\hat W_f(\tilde{w},m)e^{-2\pi\mathrm{i}mn/N},
\end{equation}
where $\hat W_f(\tilde{w},m)=\sqrt{\frac{L}{N\tilde{w}}}\hat f(m)\hat\psi(\frac{2\pi m}{N\tilde{w}})$, and $\hat\psi(k)$ is the FT of the wavelet $\psi(x)$. Clearly, as the inverse FT of the product $\hat W_f(\tilde{w},m)$, the CWT $W_f(\tilde{w},n)$ can be computed efficiently by a standard FFT routine, like the \texttt{FFTW}\footnote{\url{https://www.fftw.org/}} we used \citep{Frigo2005}.

Note that it is necessary to choose a set of discrete scales to use in Equation \eqref{eq:discrete_CWT_conv}. For the CWT, the choice of scales is arbitrary. It is convenient to discretize the scales evenly on a logarithmic scale:
\begin{equation}
\tilde{w}=\tilde{w}_\mathrm{min}2^{i+j/N_\mathrm{subs}},
\end{equation}
where $\tilde{w}_\mathrm{min}=c_w\pi/N$ is the largest scale. Here, the scales are first divided into $N_\mathrm{levs}$ levels, numbered by $i$; then each level is divided into $N_\mathrm{subs}$ sub-levels, numbered by $j$. Thus, there is a total of $N_\mathrm{scales}=N_\mathrm{levs}N_\mathrm{subs}$ scales. The number of scale levels is determined by $N_\mathrm{levs}=\mathrm{Nint}(\log_2\frac{c_wk_\mathrm{Nyq}}{\tilde{w}_\mathrm{min}/\Delta x})$, where $k_\mathrm{Nyq}$ is the Nyquist frequency and $\mathrm{Nint}(\ldots)$ denotes the nearest integer, while the number of sub-levels is determined by the user to allow adjustment of the scale resolution.

\begin{figure}
	\centerline{\includegraphics[width=0.47\textwidth]{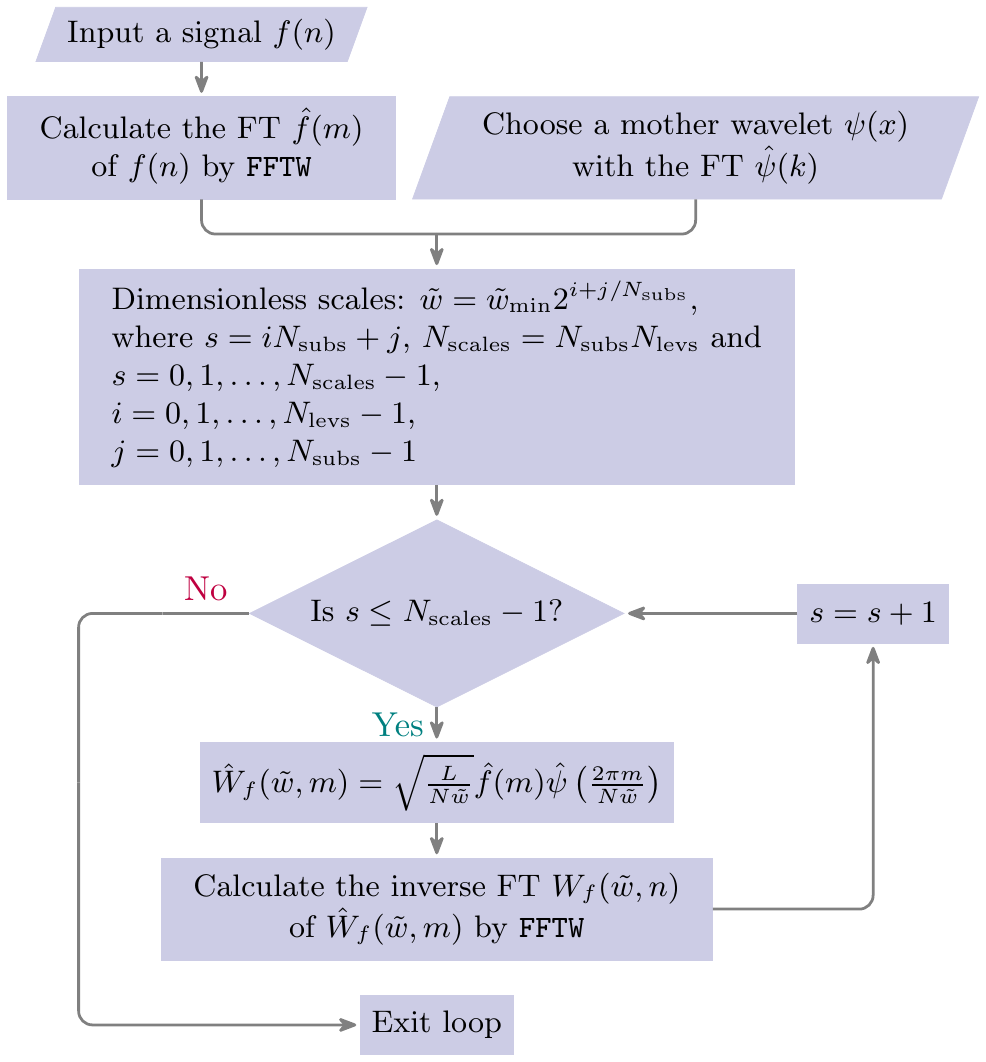}}
	\caption{Schematic representation of the FFTCWT algorithm. We define the loop variable $s=iN_\mathrm{subs}+j$ to merge the two nested loops (for $i$ and $j$) into one single loop.}
	\label{fig:fftCWT}
\end{figure}

For clarity, the sequence of the FFTCWT algorithm is shown as a flowchart in Fig. \ref{fig:fftCWT}.

\begin{figure}
	\centerline{\includegraphics[width=0.47\textwidth]{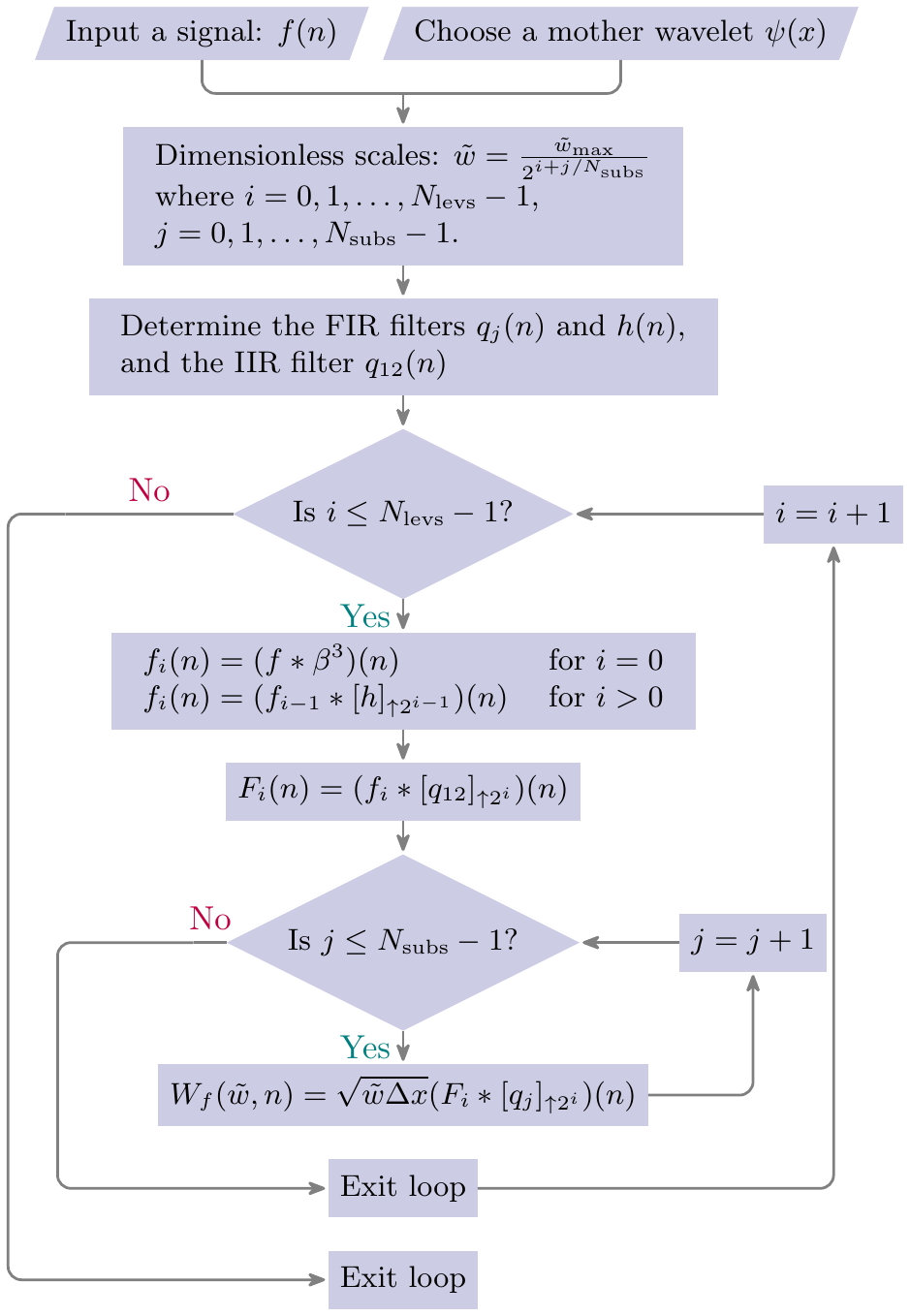}}
	\caption{Schematic representation of the V97CWT algorithm.}
	\label{fig:Vrhel97}
\end{figure}

\subsection{V97CWT}
\label{sec:v97cwt_algo}

The fundamental idea of the V97CWT algorithm is to approximate the wavelet using two scaling functions, e.g. the zero order B-spline $\beta^0(x)$ and the cubic B-spline function $\beta^3(x)$ (see Appendix \ref{sec:bsplines} for the definitions and properties of the B-splines). By using $\beta^3(x)$, the wavelet can be approximated as
\begin{align}
\label{eq:beta3_wavelet_approx}
\psi_j(\tilde{x})=\psi\left(\frac{\tilde{w}_\mathrm{max}\tilde{x}}{2^{j/N_\mathrm{subs}}}\right) &\approx (p_j*\beta^3)(\tilde{x})\nonumber\\ &=\sum_np_j(n)\beta^3(\tilde{x}-n),
\end{align}
where $\tilde{x}=x/\Delta x$ is the dimensionless coordinate, and $\tilde{w}_\mathrm{max}$ is the smallest scale. By convolving the above equation with $\beta^0(x)$, we get
\begin{equation}
\label{eq:pj}
p_j(n) = (q_j*q_{12})(n),
\end{equation}
where the FIR filter $q_j(n)$ is
\begin{align}
\label{eq:qj}
q_j(n) &= (\psi_j*\beta^0)(n)\nonumber\\
&= \int_{n{-}1/2}^{n{+}1/2}\psi_j(\tilde{x})\mathrm{d}\tilde{x}, \ \ n{=}{-}N_q,\ldots,{-1},0,1,\ldots,N_q,
\end{align}
and the IIR filter $q_{12}(n)=(\beta^4)^{-1}(n)$ is the convolution inverse of $\beta^4(n)$, i.e.
\begin{equation}
\label{eq:q_12}
(q_{12}*\beta^4)(n) = \delta^K(n),
\end{equation}
where $\delta^K(n)$ is the Kronecker delta function.

Substituting Equations \eqref{eq:beta3_wavelet_approx} and \eqref{eq:pj} into Equation \eqref{eq:discreteCWT}, we have the following equations
\begin{align}
f_0(n) &= (f*\beta^3)(n), \label{eq:fir_filter_beta}\\
F_0(n) &= (f_0*q_{12})(n), \label{eq:iir_filter_0}\\
W_f(\frac{\tilde{w}_\mathrm{max}}{2^{j/N_\mathrm{subs}}},n) &= \sqrt{\frac{\tilde{w}_\mathrm{max}\Delta x}{2^{j/N_\mathrm{subs}}}}(F_0*q_j)(n).\label{eq:fir_filter_q0}
\end{align}
Exploiting the two-scale relation of the B-splines, we can obtain the CWT at scales of $\tilde{w}_\mathrm{max}/2^{i+j/N_\mathrm{subs}}$ as follows
\begin{align}
f_i(n) &= (f_{i-1}*[h]_{\uparrow 2^{i-1}})(n), \label{eq:fir_filter_hi}\\
F_i(n) &= (f_i*[q_{12}]_{\uparrow 2^{i}})(n), \label{eq:iir_filter_i}\\
W_f(\frac{\tilde{w}_\mathrm{max}}{2^{i+j/N_\mathrm{subs}}},n) &= \sqrt{\frac{\tilde{w}_\mathrm{max}\Delta x}{2^{i+j/N_\mathrm{subs}}}}(F_i*[q_j]_{\uparrow 2^{i}})(n),\label{eq:fir_filter_qi}
\end{align}
where ``$[\ldots]_{\uparrow 2^{i}}$" denotes the insertion of $2^i-1$ zeros between each point, and $h(n)$ is given by Equation \eqref{eq:h_filter}. Equations \eqref{eq:fir_filter_beta}, \eqref{eq:fir_filter_q0}, \eqref{eq:fir_filter_hi} and \eqref{eq:fir_filter_qi} perform the FIR filtering. Equations \eqref{eq:iir_filter_0} and \eqref{eq:iir_filter_i} perform the IIR filtering, please refer to Appendix \ref{sec:iir_filter} for its details. The maximum scale level $N_\mathrm{levs}-1$ is determined by $N_\mathrm{levs}=\mathrm{Nint}(\log_2\frac{\tilde{w}_\mathrm{max}}{c_w\pi/N})$.

The sequence of the V97CWT algorithm is shown as a flowchart in Fig. \ref{fig:Vrhel97}.

\begin{figure}
	\centerline{\includegraphics[width=0.47\textwidth]{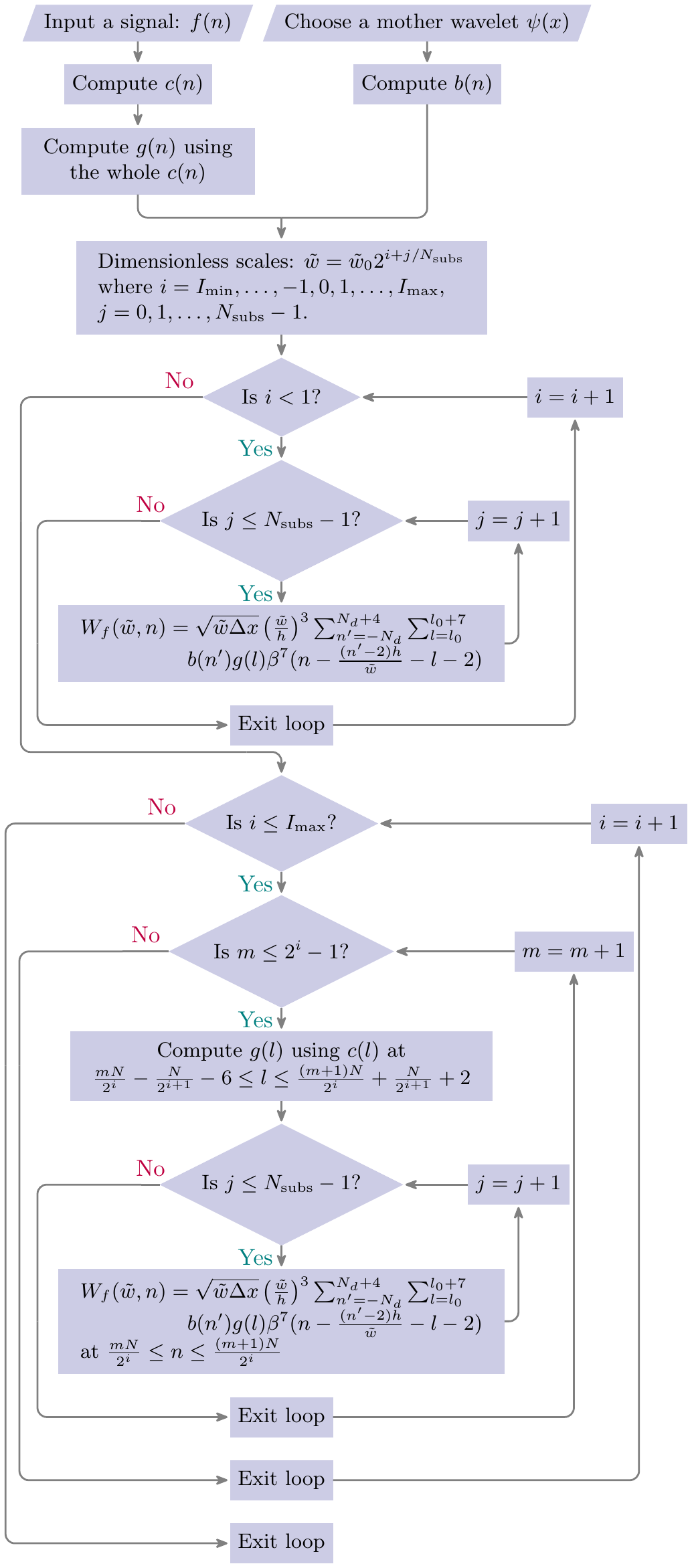}}
	\caption{Schematic representation of the M02CWT algorithm.}
	\label{fig:Munoz02}
\end{figure}

\begin{figure}
	\centerline{\includegraphics[width=0.47\textwidth]{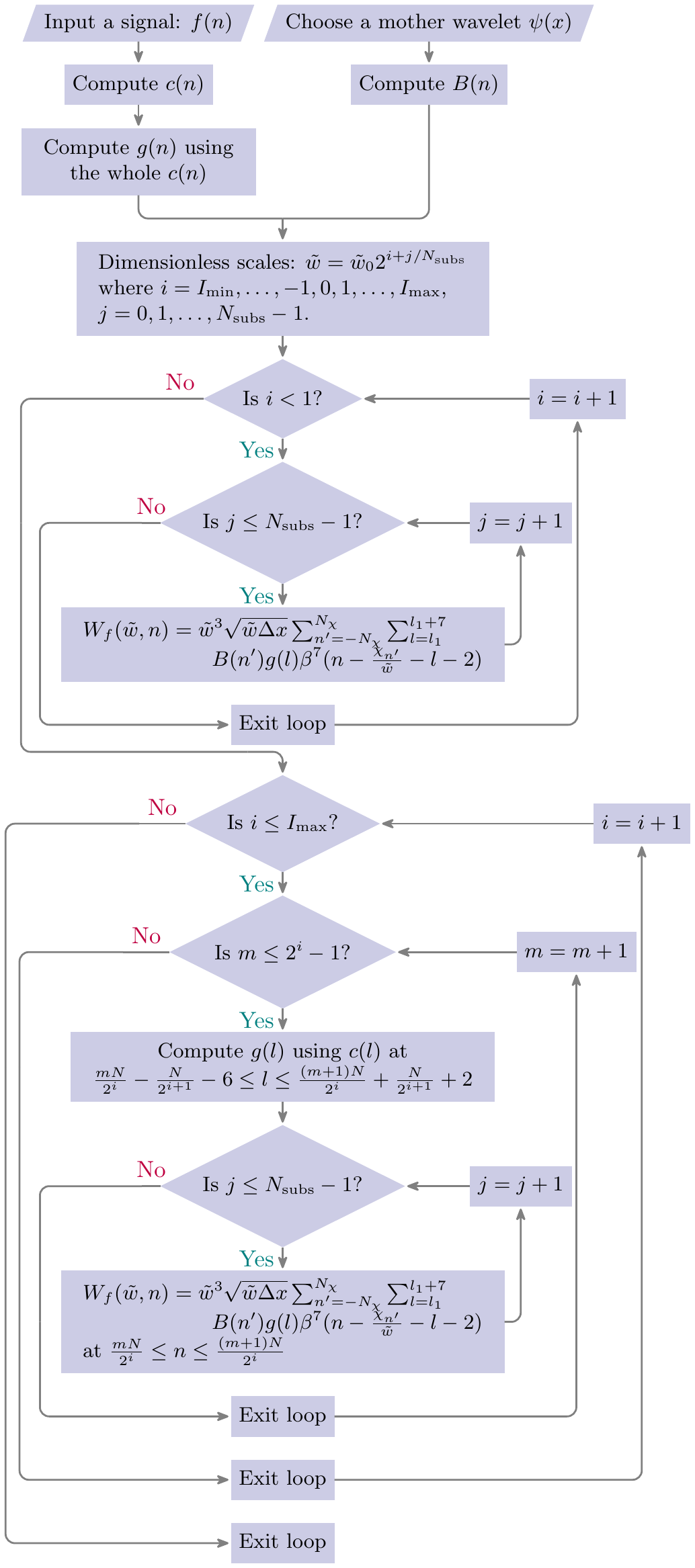}}
	\caption{Schematic representation of the A19CWT algorithm.}
	\label{fig:Arizumi19}
\end{figure}

\subsection{M02CWT}
\label{sec:m02cwt_algo}

The M02CWT algorithm represents both the wavelet and the input signal as B-splines. The wavelet function is expressed as
\begin{equation}
\label{eq:beta3_wavelet_approx_m02}
\psi(\tilde{w}\tilde{x})\approx\sum_{n=-N_d}^{N_d}d(n)\beta^3(\frac{\tilde{w}\tilde{x}}{h}-n),
\end{equation}
where the parameter $h$ is used to regulate the accuracy of the B-spline approximation. If the support of the wavelet $\psi(x)$ is $[-\chi,\chi]$, then the relation between $h$ and $N_d$ is $h=\chi/(N_d+2)$. Likewise, the continuous signal $f_c(\tilde{x})$ is represented by its cubic B-spline interpolant:
\begin{align}
\label{eq:beta3_signal}
f_c(\tilde{x}) &= (c*\beta^3)(\tilde{x})\nonumber\\
&= \sum_nc(n)\beta^3(\tilde{x}-n).
\end{align}
For spline wavelets, e.g. the CBSW, the coefficient sequence $d(n)$ can be easily obtained from its analytic form. For general wavelet functions, the sequence $d(n)$ is calculated in the same method as $c(n)$, which are calculated by (see Appendix \ref{sec:iir_filter})
\begin{equation}
\label{eq:interp_coeffs}
c(n) = \left(f*(\beta^3)^{-1}\right)(n),
\end{equation}
where $f(n)$ is the discrete input signal.

Substituting Equations \eqref{eq:beta3_wavelet_approx_m02} and \eqref{eq:rescaled_beta} into Equation \eqref{eq:CWT}, we get
\begin{equation*}
W_f(\tilde{w},\tilde{x})=\sqrt{\tilde{w}\Delta x}\left(\frac{\tilde{w}}{h} \right)^3\sum_{n=-N_d}^{N_d+4}b(n)v\left(\tilde{x}-\frac{nh}{\tilde{w}}\right),
\end{equation*}
where $b(n)=\sum_{n'=0}^{4}d(n-n')a(n')$, and $v(\tilde{x})=D^{-4}f_c(\tilde{x}+2h/\tilde{w})$. Then by using Equations \eqref{eq:beta3_signal} and \eqref{eq:antiderivative_beta}, and considering that we are typically interested in the integer values of $\tilde{x}$, the above equation becomes
\begin{align}
\label{eq:m02_cwt}
&W_f(\tilde{w},n)\nonumber= \\
&\sqrt{\tilde{w}\Delta x}\left(\frac{\tilde{w}}{h} \right)^3\sum_{n'=-N_d}^{N_d+4}\sum_{l=l_0}^{l_0+7}b(n')g(l)\beta^7\left(n{-}\frac{(n'{-}2)h}{\tilde{w}}{-}l{-}2\right),
\end{align}
where $l_0$ is the ceiling integer of $n-(n'-2)h/\tilde{w}-6$, and
\begin{equation}
\label{eq:cumsum4_coeffs}
g(l)=(\Delta^{-4}*c)(l).
\end{equation}
Notice that the computation of $g(l)$ needs to calculate cumulative sum of the sequence $c(l)$ four times, which indicates that in the case of a large amount of data, $g(l)$ becomes increasingly inaccurate as $l$ increases due to the limited precision of floating points. The way to alleviate this issue is to divide the sequence $c(l)$ into many small segments and then compute $g(l)$ locally on each segment. To do this, we set scales as follows
\begin{equation}
\tilde{w}=\tilde{w}_02^{i+j/N_\mathrm{subs}},
\end{equation}
where $\tilde{w}_0=2\chi/N$, and the scale level $i$ takes the range of $I_\mathrm{min}=\mathrm{Nint}(\log_2\frac{c_w\pi/N}{\tilde{w}_0})$ to $I_\mathrm{max}=\mathrm{Nint}(\log_2\frac{c_w\pi/2}{\tilde{w}_0})$. In the case of $i<1$, we do not split $c(l)$; whereas in the case of $i\geq1$, we split $c(l)$ into $2^i$ parts, as shown by the flowchart in Fig. \ref{fig:Munoz02}.

\subsection{A19CWT}
\label{sec:a19cwt_algo}

Since the mother wavelet is zero outside the support interval $[-\chi, \chi]$, the CWT at integer positions can be written as
\begin{equation}
\label{eq:a19cwt_01}
W_f(\tilde{w},n) = \sqrt{\frac{\Delta x}{\tilde{w}}}\int_{-\chi}^{\chi}f_c(n-\tilde{u}/\tilde{w})\psi(\tilde{u})\mathrm{d}\tilde{u}.
\end{equation}
By partitioning the support duration $[-\chi, \chi]$ evenly into $2N_\chi$ intervals, i.e.
\begin{equation*}
-\chi=\chi_{-N_\chi}<\ldots<\chi_{-1}<\chi_{0}<\chi_{1}<\ldots<\chi_{N_\chi}=\chi,
\end{equation*}
we approximate $\psi(\tilde{x})$ with cubic piecewise polynomials as shown below
\begin{equation}
\label{eq:piecewise_polys_psi}
\psi(\tilde{x})\approx
\begin{cases}
\psi_{{-}N_\chi}(\tilde{x}), & \chi_{{-}N_\chi}\leq \tilde{x}<\chi_{1{-}N_\chi},\\
\vdots & \vdots\\
\psi_{n'}(\tilde{x}), & \chi_{n'}\leq \tilde{x}<\chi_{n'+1},\\
\vdots & \vdots\\
\psi_{N_\chi{-}1}(\tilde{x}), & \chi_{N_\chi{-}1}\leq \tilde{x}<\chi_{N_\chi},\\
\end{cases}
\end{equation}
where $\psi_{n'}(\tilde{x})=\sum_{i=0}^3\alpha_{n',i}(\tilde{x}-\chi_{n'})^i$ for $\chi_{n'}\leq \tilde{x}<\chi_{n'+1}$.

Substituting Equation \eqref{eq:piecewise_polys_psi} into Equation \eqref{eq:a19cwt_01}, we have
\begin{equation}
\label{eq:a19cwt_02}
W_f(\tilde{w},n) = \sqrt{\frac{\Delta x}{\tilde{w}}}{\sum_{n'=-N_\chi}^{N_\chi-1}}\int_{\chi_{n'}}^{\chi_{n'{+}1}}f_c\left(n{-}\frac{\tilde{u}}{\tilde{w}}\right)\psi_{n'}(\tilde{u})\mathrm{d}\tilde{u}.
\end{equation}
Then we apply integration by parts to Equation \eqref{eq:a19cwt_02} and arrive at
\begin{align}
\label{eq:a19cwt_03}
W_f(\tilde{w},n) &= \tilde{w}^3\sqrt{\frac{\Delta x}{\tilde{w}}}{\sum_{n'=-N_\chi}^{N_\chi-1}}{6\alpha_{n',3}}{\int_{\chi_{n'}}^{\chi_{n'+1}}}F_c^{(3)}\left(n{-}\frac{\tilde{u}}{\tilde{w}}\right)\mathrm{d}\tilde{u}\nonumber\\
&= \tilde{w}^4\sqrt{\frac{\Delta x}{\tilde{w}}}{\sum_{n'=-N_\chi}^{N_\chi-1}}{6\alpha_{n',3}}\Big(F_c^{(4)}(n{-}\frac{\chi_{n'}}{\tilde{w}})\nonumber\\
&\qquad\qquad\qquad\qquad\qquad\quad-F_c^{(4)}(n{-}\frac{\chi_{n'+1}}{\tilde{w}})\Big)\nonumber\\
&=\tilde{w}^3\sqrt{\tilde{w}\Delta x}{\sum_{n'=-N_\chi}^{N_\chi}}B(n')F_c^{(4)}(n{-}\frac{\chi_{n'}}{\tilde{w}}),
\end{align}
where $F_c^{(4)}(\tilde{x})$ is the 4th antiderivative of $f_c(\tilde{x})$, and
\begin{align}
B(-N_\chi) &= 6\alpha_{-N_\chi,3},\nonumber\\
B(n) &= 6(\alpha_{n,3}-\alpha_{n-1,3}), \quad\mathrm{for}\quad 1-N_\chi\leq n\leq N_\chi-1,\nonumber\\
B(N_\chi) &= -6\alpha_{N_\chi-1}.
\end{align}
In fact, Equation \eqref{eq:a19cwt_03} assumes that the 3rd derivative of the wavelet, i.e. $\psi'''$ is constant on the interval $[\chi_{n'},\chi_{n'+1})$, which can be approximated as
\begin{equation}
\label{eq:3rd_derivative_wavelet}
\psi'''(\tilde{x}) \approx\frac{\psi''(\chi_{n'+1})-\psi''(\chi_{n'})}{\chi_{n'+1}-\chi_{n'}},
\end{equation}
where $\psi''$ is the 2nd derivative of the wavelet, which can be obtained analytically. Hence the coefficient $\alpha_{n',3}$ is given by
\begin{equation}
\alpha_{n',3} = \frac{\psi''(\chi_{n'+1})-\psi''(\chi_{n'})}{6(\chi_{n'+1}-\chi_{n'})}.
\end{equation}

By using Equations \eqref{eq:beta3_signal}, \eqref{eq:integral_operator} and \eqref{eq:antiderivative_beta}, $F_c^{(4)}(\tilde{x})$ can be calculated as
\begin{equation}
\label{eq:4th_antiderivative}
F_c^{(4)}(\tilde{x})=\sum_lg(l)\beta^7(\tilde{x}-l-2).
\end{equation}
Therefore Equation \eqref{eq:a19cwt_03} can be expressed as
\begin{align}
\label{eq:a19cwt_04}
& W_f(\tilde{w},n) = \nonumber\\
&\qquad\tilde{w}^3\sqrt{\tilde{w}\Delta x}{\sum_{n'=-N_\chi}^{N_\chi}}{\sum_{l=l_1}^{l_1+7}}B(n')g(l)\beta^7(n{-}\frac{\chi_{n'}}{\tilde{w}}{-}l{-}2),
\end{align}
where $l_1$ is the ceiling integer of $n-\chi_{n'}/\tilde{w}-6$, and the coefficient sequence $g(l)$ is computed by Equation \eqref{eq:cumsum4_coeffs}. By comparing Equations \eqref{eq:m02_cwt} and \eqref{eq:a19cwt_04}, we find that the M02CWT and A19CWT are very similar, but the theoretical derivation of the A19CWT is much simpler. To solve the accuracy issue of $g(l)$, we adopt the same scheme as the M02CWT algorithm.

The sequence of the A19CWT algorithm is shown as a flowchart in Fig. \ref{fig:Arizumi19}.

\begin{figure}
	\centerline{\includegraphics[width=0.48\textwidth]{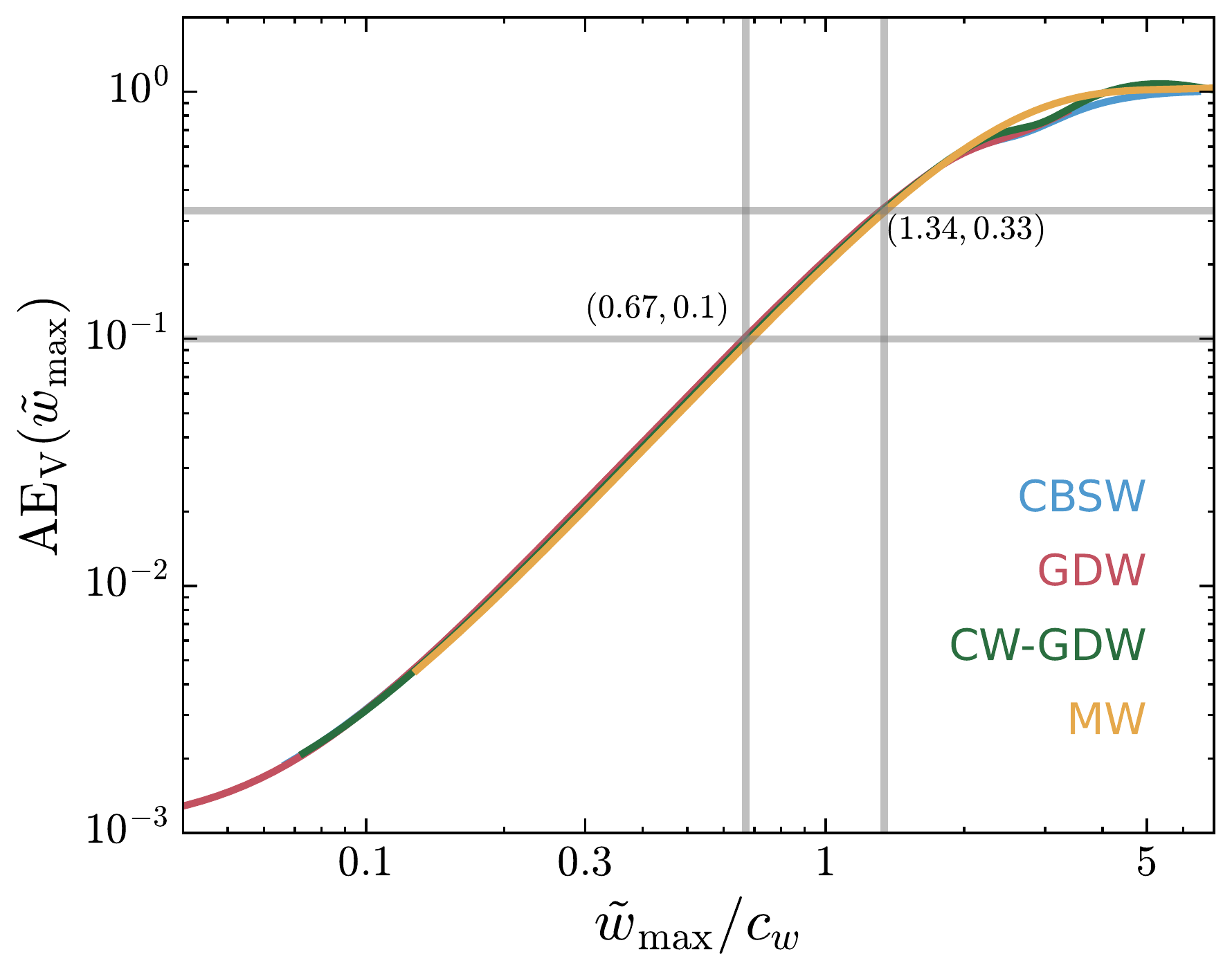}}
	\caption{The Approximation error defined by Equation \eqref{eq:v97cwt_error} for different wavelets as labeled. At $\tilde{w}_\mathrm{max}=0.67c_w$, the error reaches $0.1$. When $\tilde{w}_\mathrm{max}=1.34c_w$, i.e., twice as large as $0.67c_w$, the error is roughly $0.33$.}
	\label{fig:approx_err_v97cwt}
\end{figure}

\begin{figure}
	\centerline{\includegraphics[width=0.48\textwidth]{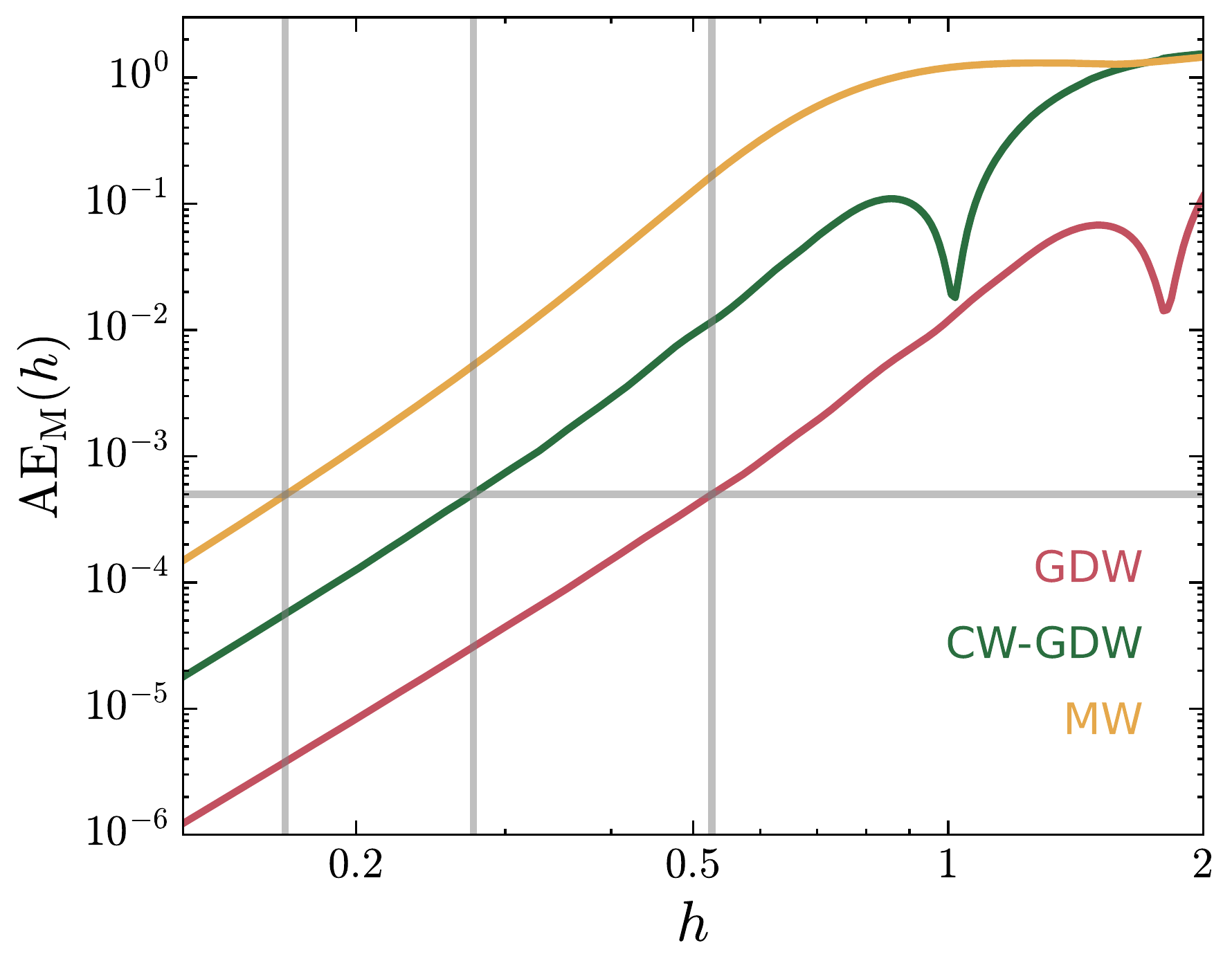}}
	\caption{The Approximation error defined by Equation \eqref{eq:m02cwt_error} for different wavelets as labeled. The gray horizontal line shows the error level of $5\times 10^{-4}$, and the gray vertical lines denote $h$'s values where the error reaches $5\times 10^{-4}$ for the MW, CW-GDW and GDW from left to right.}
	\label{fig:approx_err_m02cwt}
\end{figure}

\begin{figure}
	\centerline{\includegraphics[width=0.48\textwidth]{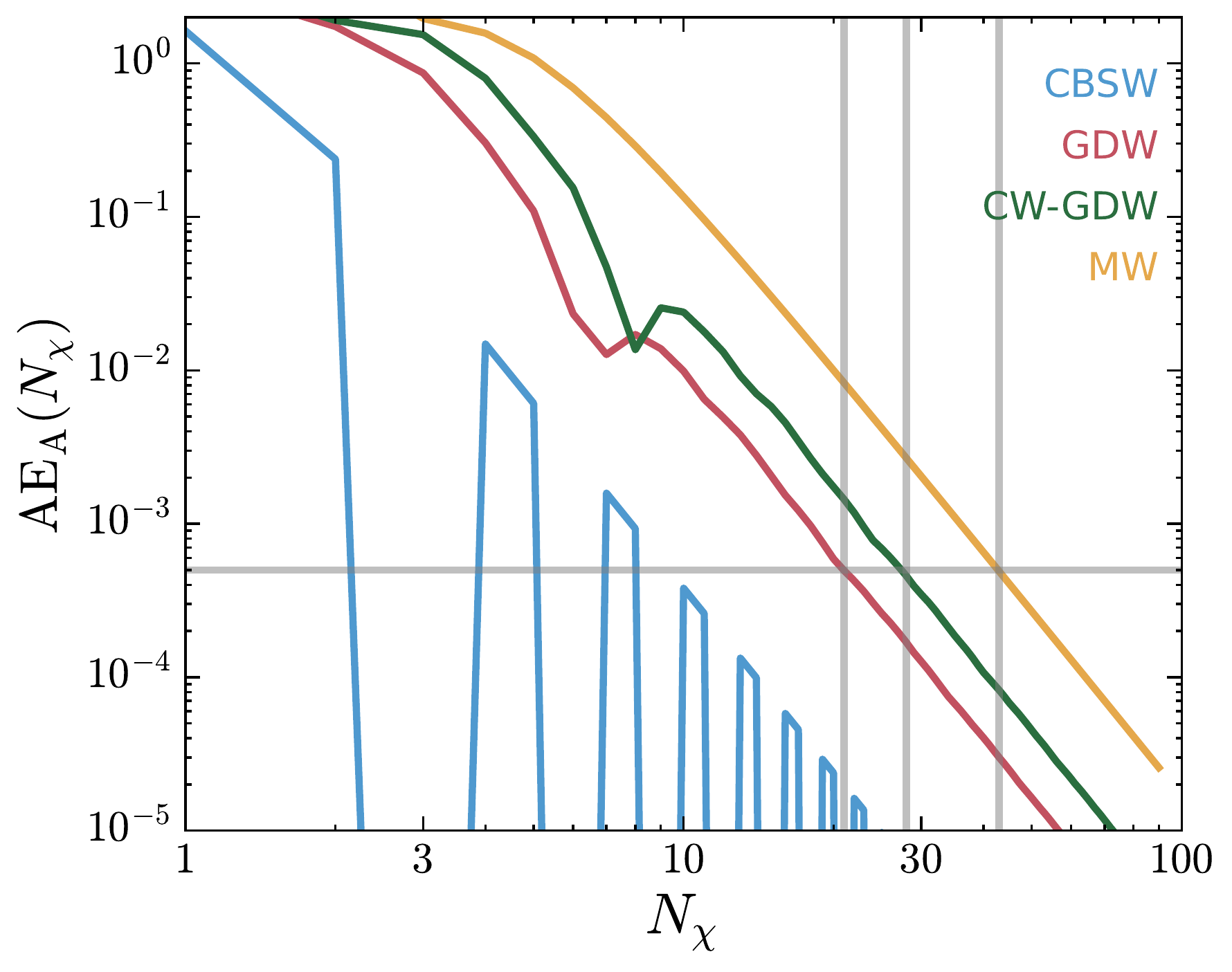}}
	\caption{The Approximation error defined by Equation \eqref{eq:a19cwt_error} for different wavelets as labeled. For the CBSW, at $N_\chi$ equal to the integer multiples of $3$, the error is very tiny and nearly $10^{-16}$. So $N_\chi=3$ is the best choice for it. The gray horizontal line shows the error level of $5\times 10^{-4}$, and the gray vertical lines denote $N_\chi$'s values where the error reaches $5\times 10^{-4}$ for the GDW, CW-GDW and MW from left to right.}
	\label{fig:approx_err_a19cwt}
\end{figure}

\subsection{Boundary conditions}

In the definition of the CWT (see Equation \eqref{eq:CWT}), the signal is assumed to be extended to infinity. Nevertheless, in reality, the length of the analyzed signal is finite. Hence, we must make assumptions about the data outside its finite extent. Periodic boundary conditions are the most common choice. On the one hand, many cosmic fields are considered to be periodic. On the other hand, periodic boundary conditions are easy to implement. The FFTCWT inherits the attribute that the signal is assumed to be periodic in the FFT. The V97CWT imposes periodic boundary conditions on the signal by the IIR filtering (see Appendix \ref{sec:iir_filter}). For the M02CWT and A19CWT, the fact that the signal is periodic only imply that the coefficient $c(l)$ is periodic. To ensure that $g(l)$ is periodic, we take the following operations:

\begin{algorithm}
	\begin{algorithmic}[1]
		\If{$i<1$}   \Comment{\textcolor{gray}{Scale levels less than 1}}
		\State $g(0:N-1)\gets c(0:N-1)$
		\For{$j=1$ to $4$}
		\State $g\gets\Delta^{-1}*g$
		\State $g\gets g-\texttt{Mean}(g)$
		\EndFor
		\ElsIf{$i\geq1$}  \Comment{\textcolor{gray}{Scale levels greater than or equal to 1}}
		\State Periodically padding $\frac{N}{4}{+}6$ values at start and $\frac{N}{4}{+}3$ values at end of $c$
		\For{$m=0$ to $2^i-1$}
		\State $l_{t1}\gets \frac{mN}{2^i}-\frac{N}{2^{i+1}}-6$, $l_{t2}\gets \frac{(m+1)N}{2^i}+\frac{N}{2^{i+1}}+2$
		\State $g(l_{t1}:l_{t2})\gets(\Delta^{-4}*c)(l_{t1}:l_{t2})$
		\EndFor
		\EndIf
	\end{algorithmic}
	\label{alg:periodic_gl}
\end{algorithm}

In addition, the M02CWT and A19CWT can also easily handle signals with zero boundaries without padding zeros at scale levels of $i<1$. After calculating $c(l)$ by Equations \eqref{eq:iir_filter_real_a}, \eqref{eq:iir_filter_real_b}, and \eqref{eq:zero_initial_a}-\eqref{eq:zero_initial_d}, the coefficient $g(l)$ of  the signal with zero boundaries can be obtained by the following operations:

\begin{algorithm}[h]
	\begin{algorithmic}[1]
		\If{$i<1$}   \Comment{\textcolor{gray}{Scale levels less than 1}}
		\State $g(-6:N+1)\gets c(-6:N+1)$
		\For{$i=1$ to $4$}
		\State $g\gets\Delta^{-1}*g$
		\State $C_i\gets g(N+1)$
		\EndFor
		\For{$l=l_0$ to $l_0+7$} \Comment{\textcolor{gray}{Replace $l_0$ with $l_1$ in the A19CWT}}
		\If{$l<-6$}
		\State $g(l) \gets 0$
		\ElsIf{$l>N+1$}
		\State $l'\gets l-(N+1)$
		\State $g(l) \gets C_4{+}C_3l'{+}\frac{1}{2}C_2l'(l'{+}1){+}\frac{1}{6}C_1l'(l'{+}1)(l'{+}2)$
		\EndIf
		\EndFor
		\ElsIf{$i\geq1$}  \Comment{\textcolor{gray}{Scale levels greater than or equal to 1}}
		\State Padding $\frac{N}{4}$ zeros at start and $\frac{N}{4}{+}1$ zeros at end of $c$
		\For{$m=0$ to $2^i-1$}
		\State $l_{t1}\gets \frac{mN}{2^i}-\frac{N}{2^{i+1}}-6$, $l_{t2}\gets \frac{(m+1)N}{2^i}+\frac{N}{2^{i+1}}+2$
		\State $g(l_{t1}:l_{t2})\gets(\Delta^{-4}*c)(l_{t1}:l_{t2})$
		\EndFor
		\EndIf
	\end{algorithmic}
	\label{alg:zero_gl}
\end{algorithm}

\subsection{Parameter settings}

\begin{table}
	\centering
	\caption{Parameter settings for the V97CWT, M02CWT, and A19CWT algorithms.}
	\label{tab:para_settings}
	\noindent
	\resizebox{0.5\textwidth}{!}{
		\begin{tabular}{@{}cccccc}
			\hline
			& $\tilde{w}_\mathrm{max}$ & $N_q$ & $h$ & $N_d$ & $N_\chi$\\ \\[-1.2em]
			\hline
			CBSW   & $0.62457\ ^a$ & $9$  & $1$     & $1$  & $3$  \\ \\[-0.5em]
			GDW    & $1.19853\ ^b$ & $20$ & $0.526$  & $21$ & $21$ \\ \\[-0.5em]
			CW-GDW & $0.57381\ ^c$ & $27$ & $0.275$  & $27$ & $28$\\ \\[-0.5em]
			MW     & $0.32514\ ^d$ & $46$ & $0.165$ & $43$ & $43$ \\ \\[-0.5em]
			\hline
			\multicolumn{4}{l}{$^a$ $0.6245669798462486$}\\
			\multicolumn{4}{l}{$^b$ $1.1985324359398872$}\\
			\multicolumn{4}{l}{$^c$ $0.5738133082169298$}\\
			\multicolumn{4}{l}{$^d$ $0.3251384995061764$}
		\end{tabular}
	}
\end{table}

There are some unspecified parameters in the above algorithms, which are $\tilde{w}_\mathrm{max}$, $N_q$, $h$, $N_d$, and $N_\chi$. In this subsection, we will discuss how to tune these parameters to make the algorithms sufficiently precise and efficient.

For the V97CWT algorithm, we define the approximation error as below
\begin{equation}
\label{eq:v97cwt_error}
\mathrm{AE_V}(\tilde{w}_\mathrm{max})=\frac{\sum_n\left|\psi(\tilde{w}_\mathrm{max}\tilde{x}_n)-\sum_mp(m)\beta^3(\tilde{x}_n{-}m)\right|}{\sum_n|\psi(\tilde{w}_\mathrm{max}\tilde{x}_n)|},
\end{equation}
the result of which is shown in Fig. \ref{fig:approx_err_v97cwt}. We find that the error $\mathrm{AE_V}(\tilde{w}_\mathrm{max})$ increases with increasing $\tilde{w}_\mathrm{max}$. To ensure a high precision as well as a sufficiently large scale range, we set $\tilde{w}_\mathrm{max}=1.34c_w$, which satisfy $\mathrm{AE_V}(\tilde{w}_\mathrm{max}/2)=0.1$. According to Equation \eqref{eq:qj}, $q_j(N_q+1)=0$ yields the relationship $N_q(j) = 2^{j/N_\mathrm{subs}}\chi/\tilde{w}_\mathrm{max}-1/2$. For simplicity, we use the same value of $N_q(j)$ at each $j$ level, i.e.
\begin{equation}
\label{eq:wmax_nq_relation}
N_q = \frac{2\chi}{\tilde{w}_\mathrm{max}}-\frac{1}{2},
\end{equation}
which is the upper limit of $N_q(j)$.

For the M02CWT algorithm, we define the approximation error as below
\begin{equation}
\label{eq:m02cwt_error}
\mathrm{AE_M}(h)=\frac{\sum_n\left|\psi(\tilde{x}_n)-\sum_md(m)\beta^3(\tilde{x}_n/h{-}m)\right|}{\sum_n|\psi(\tilde{x}_n)|},
\end{equation}
the result of which is shown in Fig. \ref{fig:approx_err_m02cwt}. Since the cubic spline decomposition is perfectly exact to represent the CBSW with $h=1$ (see Table \ref{tab:wavelets}), we only consider the approximation error for the GDW, CW-GDW and MW. We see that the smaller the $h$, the smaller the error. However, considering the efficiency of the algorithm, the value of $h$ cannot be chosen too small. Hence, we set the value of $h$ such that the error $\mathrm{AE_M}(h)$ equals $5\times10^{-4}$, and then $N_d$ can be determined by $h=\chi/(N_d+2)$.

For the A19CWT algorithm, we define the approximation error as below
\begin{equation}
\label{eq:a19cwt_error}
\mathrm{AE_A}(N_\chi)=\frac{\sum_n\left|\psi(\tilde{x}_n)-\psi_\mathrm{pp}(\tilde{x}_n)\right|}{\sum_n|\psi(\tilde{x}_n)|},
\end{equation}
where $\psi_\mathrm{pp}(\tilde{x}_n)$ is the piecewise polynomial function given by Equation \eqref{eq:piecewise_polys_psi}. Because the CBSW is actually a cubic piecewise polynomial function with compact support width $2\chi=6$ and segment width $\Delta\chi=1$, the approximation error $\mathrm{AE_A}(N_\chi)$ should be very small at the integer multiples of $3$, which is illustrated in Fig. \ref{fig:approx_err_a19cwt}. Hence for the CBSW, $N_\chi=3$ is the best choice. For the GDW, CW-GDW and MW, we set the value of $N_\chi$ such that the error $\mathrm{AE_A}(N_\chi)$ roughly equals $5\times10^{-4}$, which is in accordance with the parameter settings of the M02CWT.

For clarity and convenience, we list the parameters and their values in Table \ref{tab:para_settings}.

\begin{figure*}
	\centerline{\includegraphics[width=0.9\textwidth]{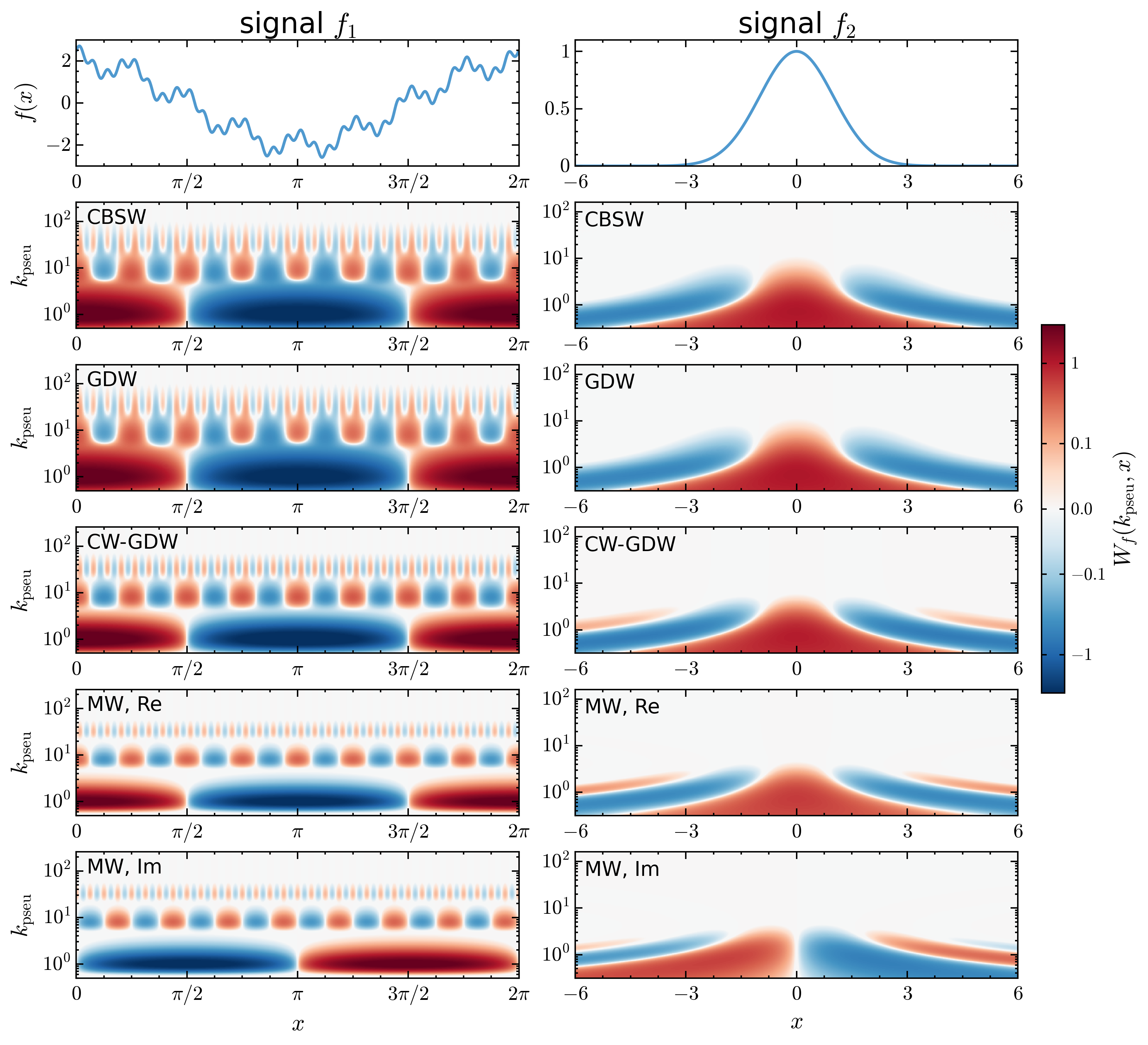}}
	\caption{\textit{Left column}: the periodic signal $f_1$ and its CWTs which are calculated analytically for different wavelets. \textit{Right column}: the same as the left column but for the non-periodic signal $f_2$. For comparison between different wavelets, we replace the wavelet scale $w$ by the pseudo wavenumber $k_\mathrm{pseu}$ (see Table \ref{tab:wavelets}), and keep this convention throughout the subsequent plots.}
	\label{fig:test_signals_cwts}
\end{figure*}

\begin{figure*}
	\centering
	\subfigure{\includegraphics[width=0.44\textwidth]{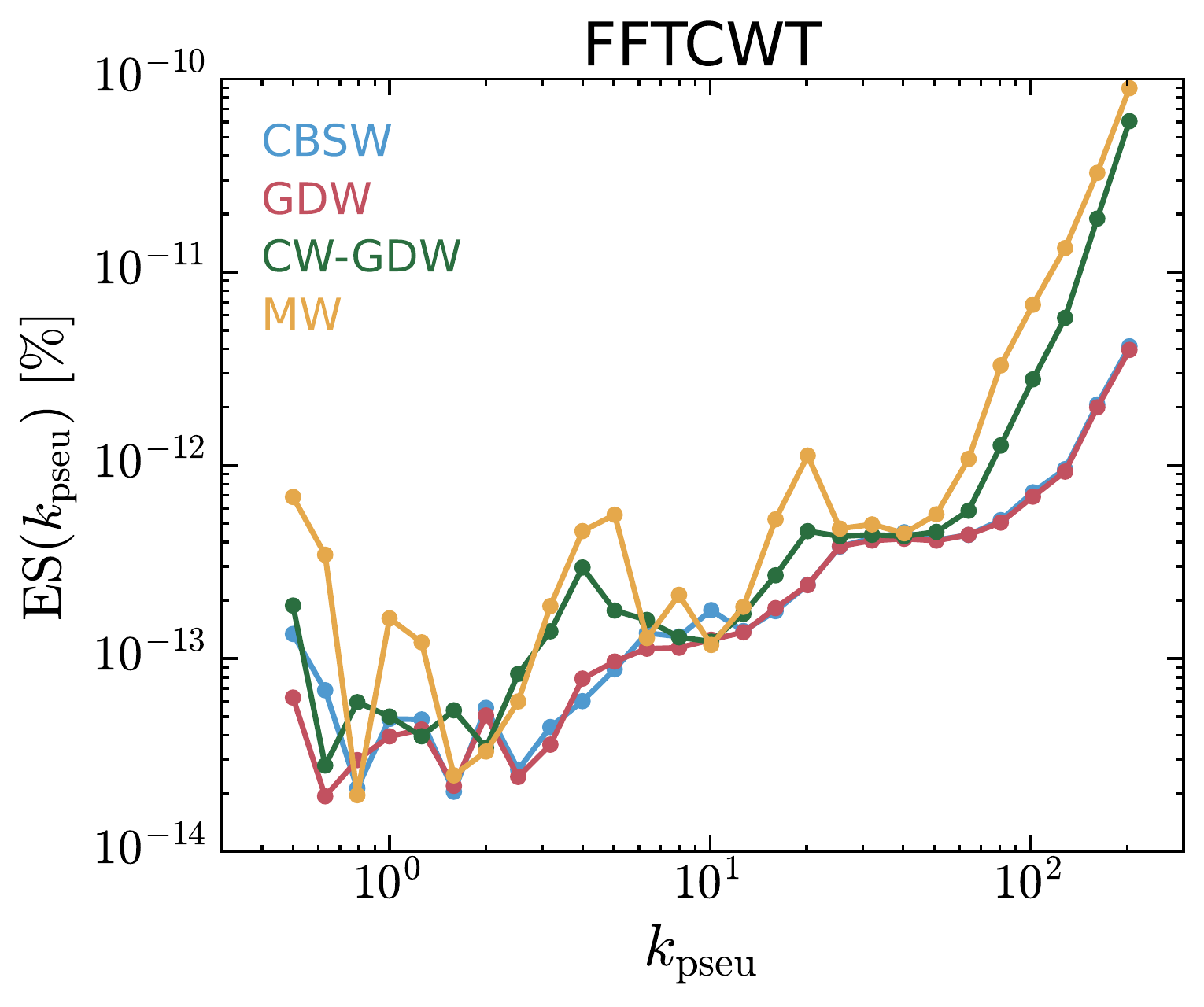}}
	\subfigure{\includegraphics[width=0.44\textwidth]{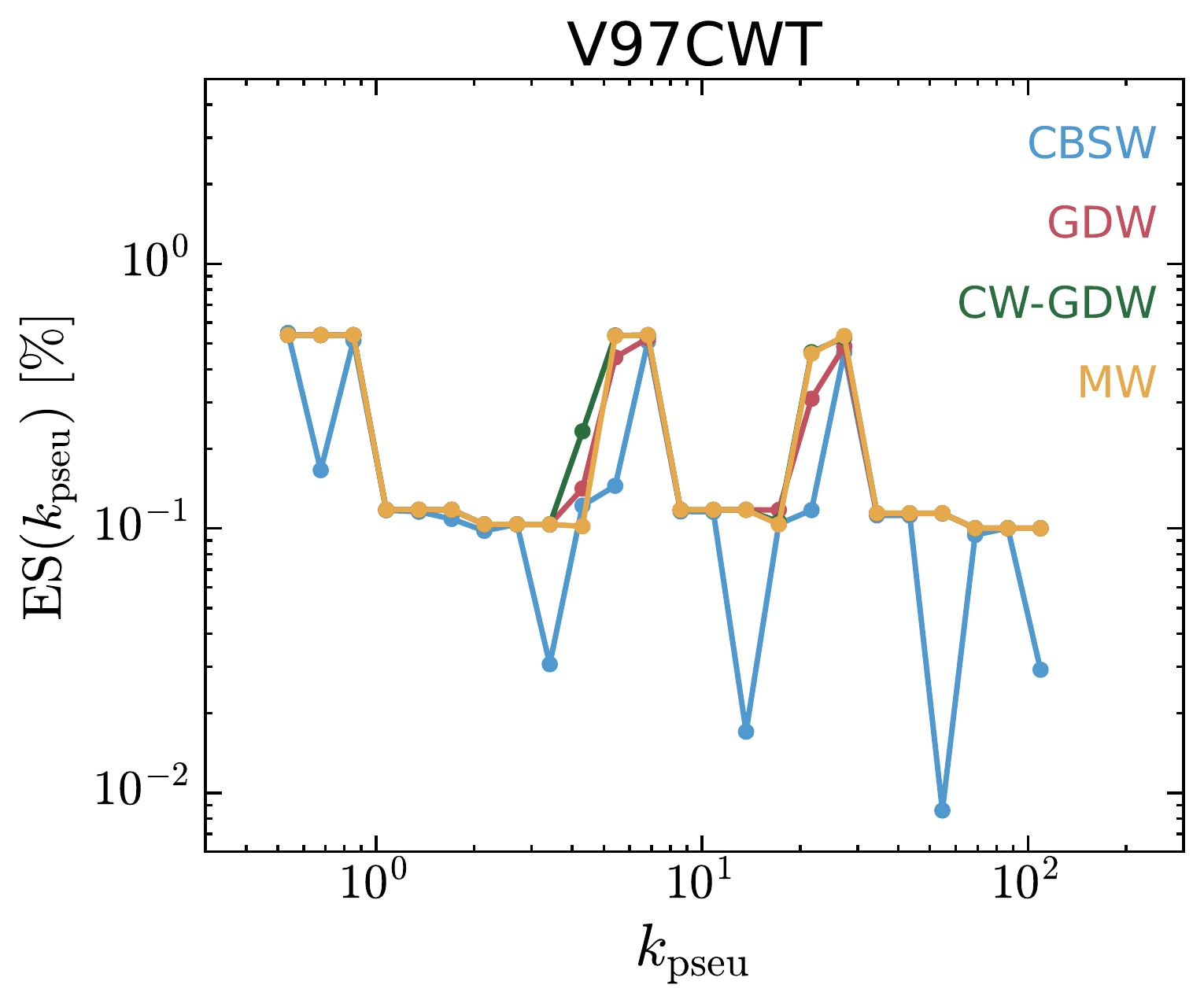}}
	\subfigure{\includegraphics[width=0.44\textwidth]{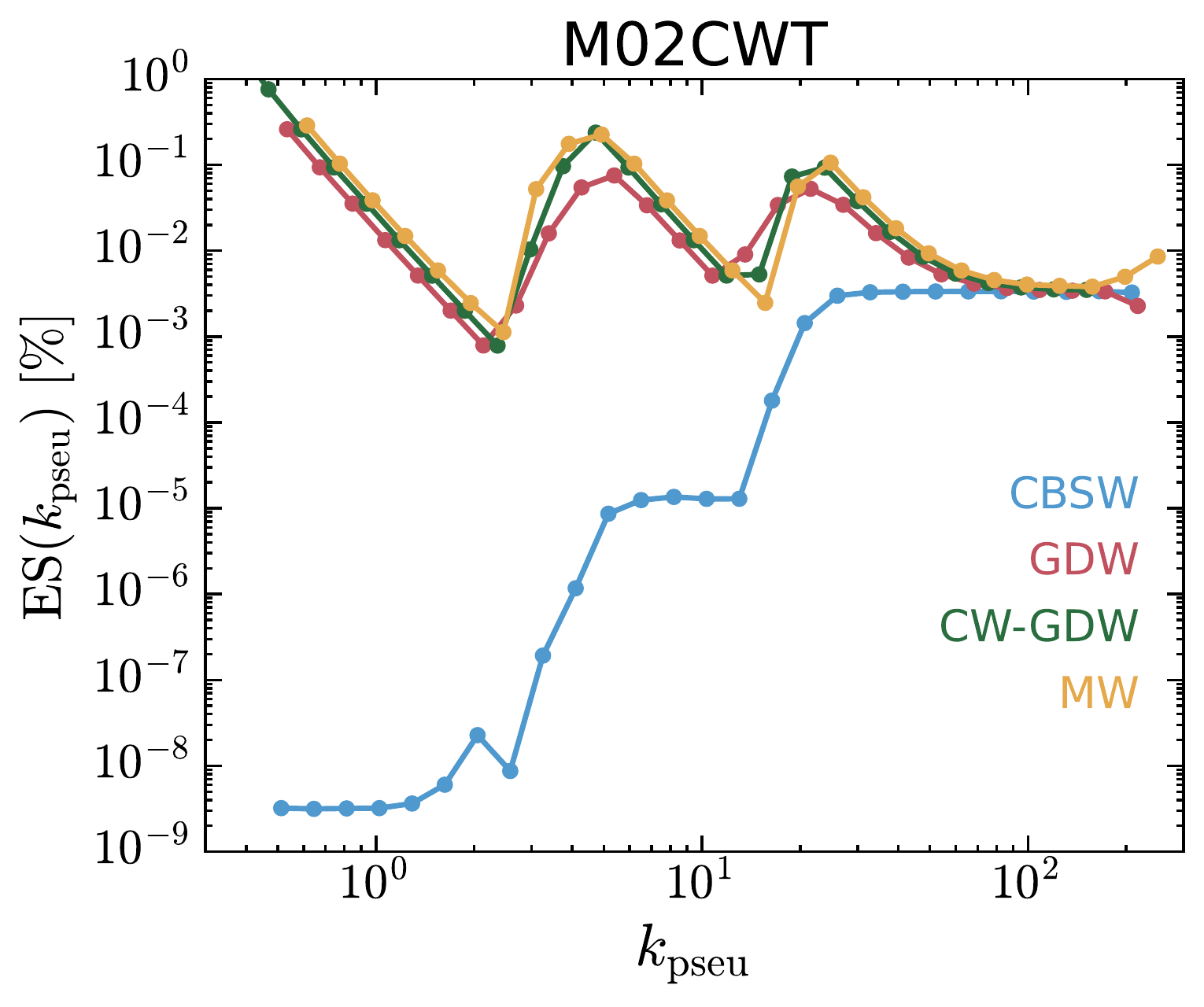}}
	\subfigure{\includegraphics[width=0.44\textwidth]{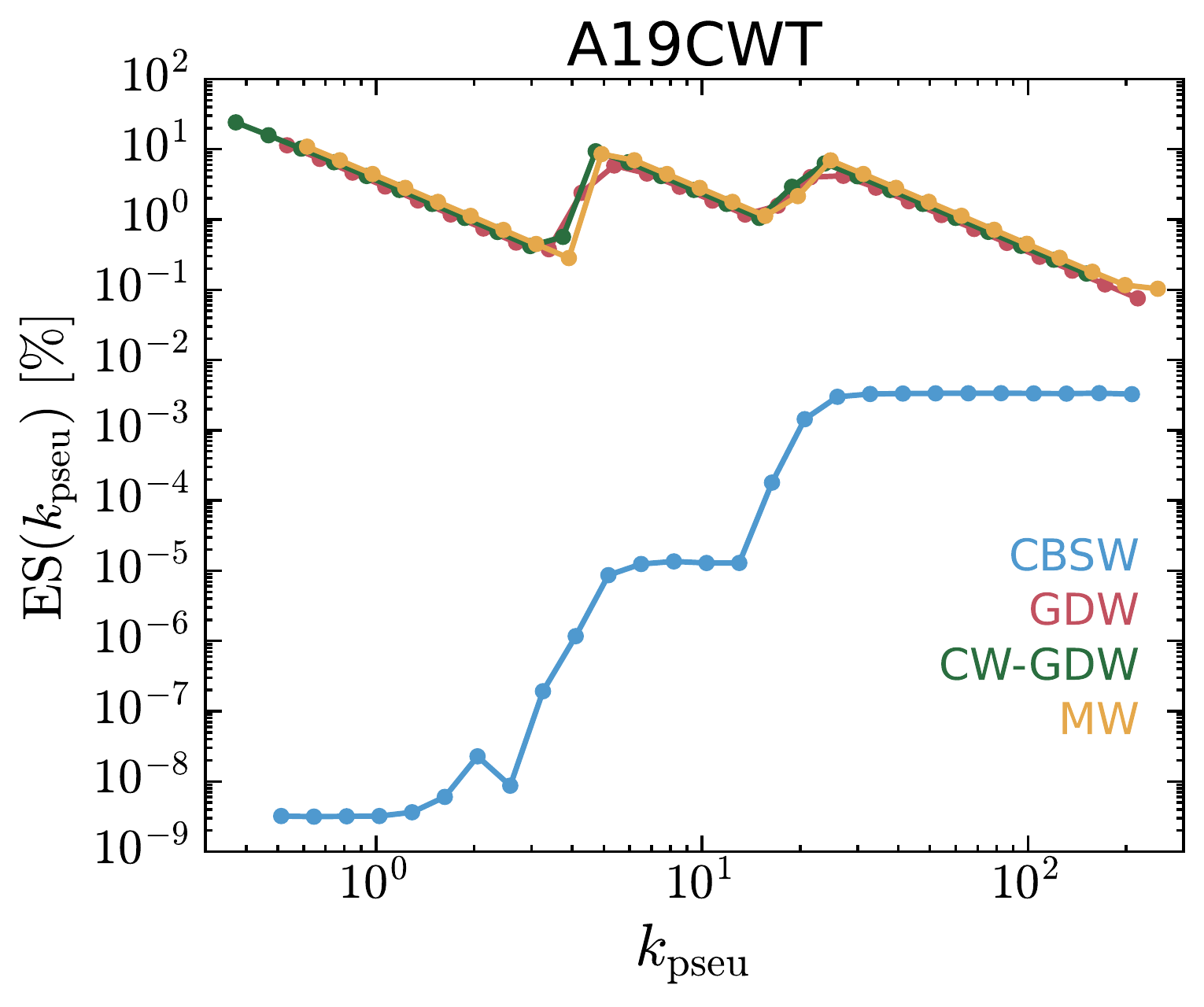}}
	\caption{The error spectra of the periodic signal $f_1(x)$ for different wavelets, computed by the FFTCWT, V97CWT, M02CWT and A19CWT algorithms, respectively.}
	\label{fig:es_spectra_periodic}
\end{figure*}

\begin{figure*}
	\centering
	\subfigure{\includegraphics[width=0.44\textwidth]{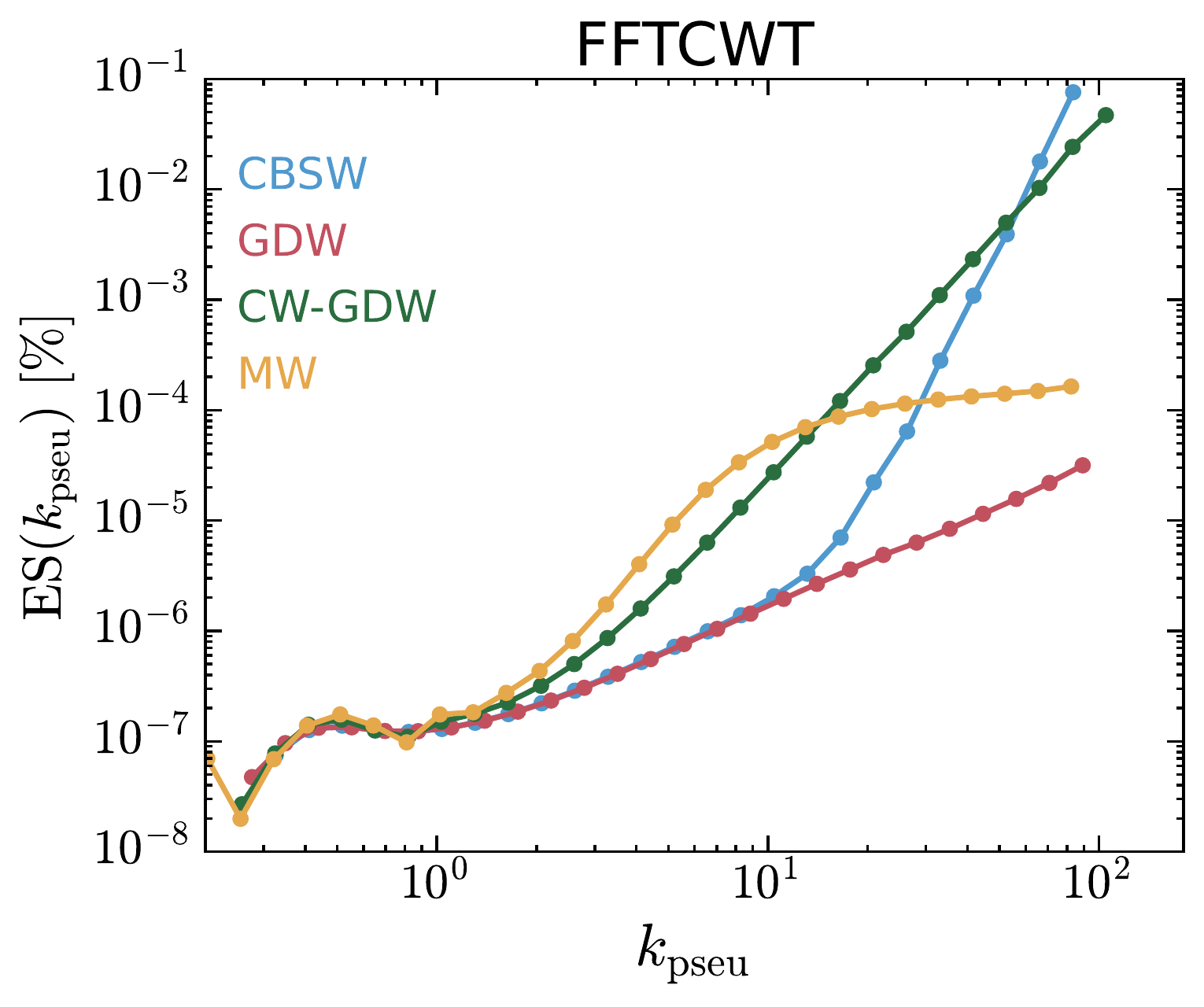}}
	\subfigure{\includegraphics[width=0.44\textwidth]{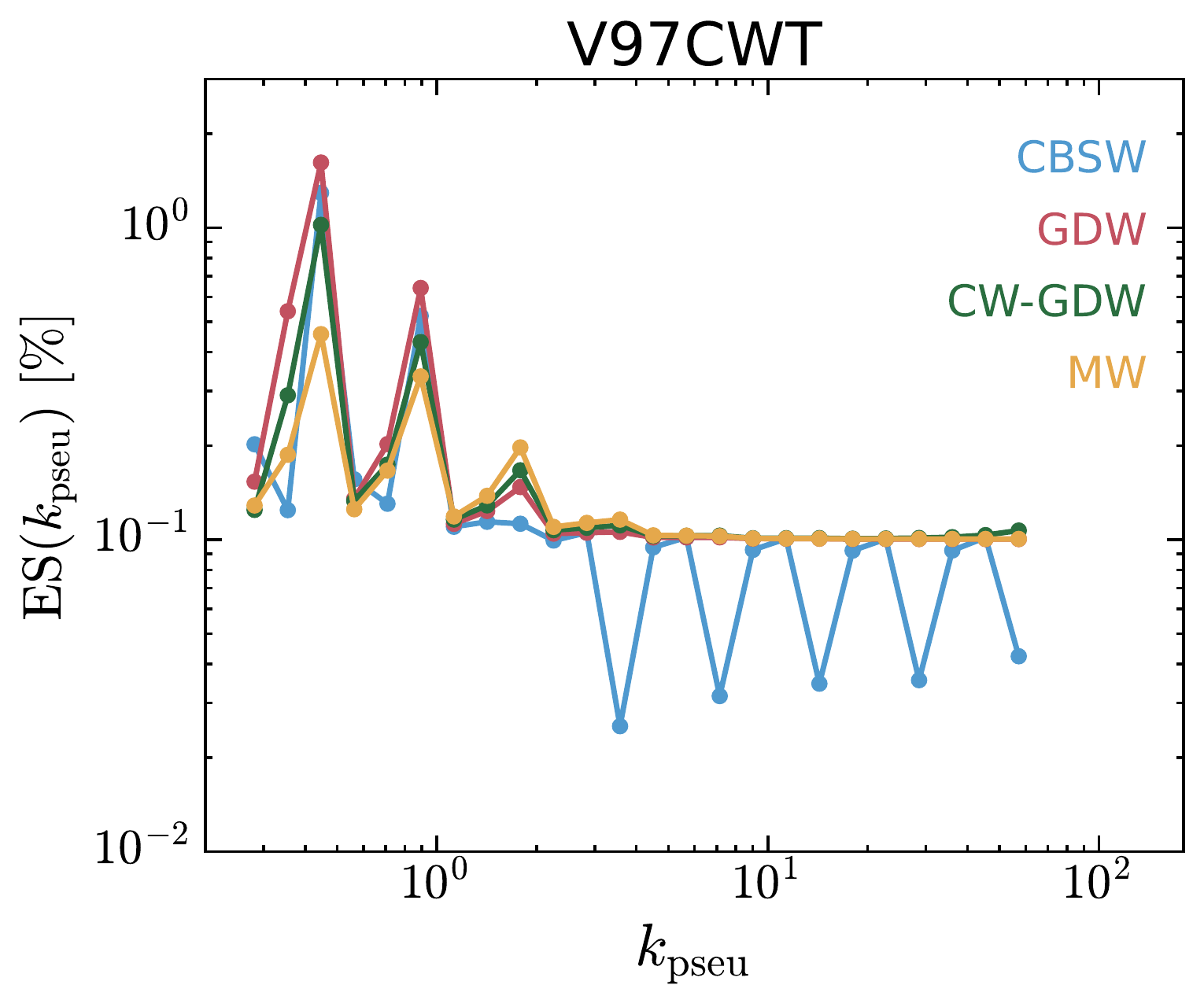}}
	\subfigure{\includegraphics[width=0.44\textwidth]{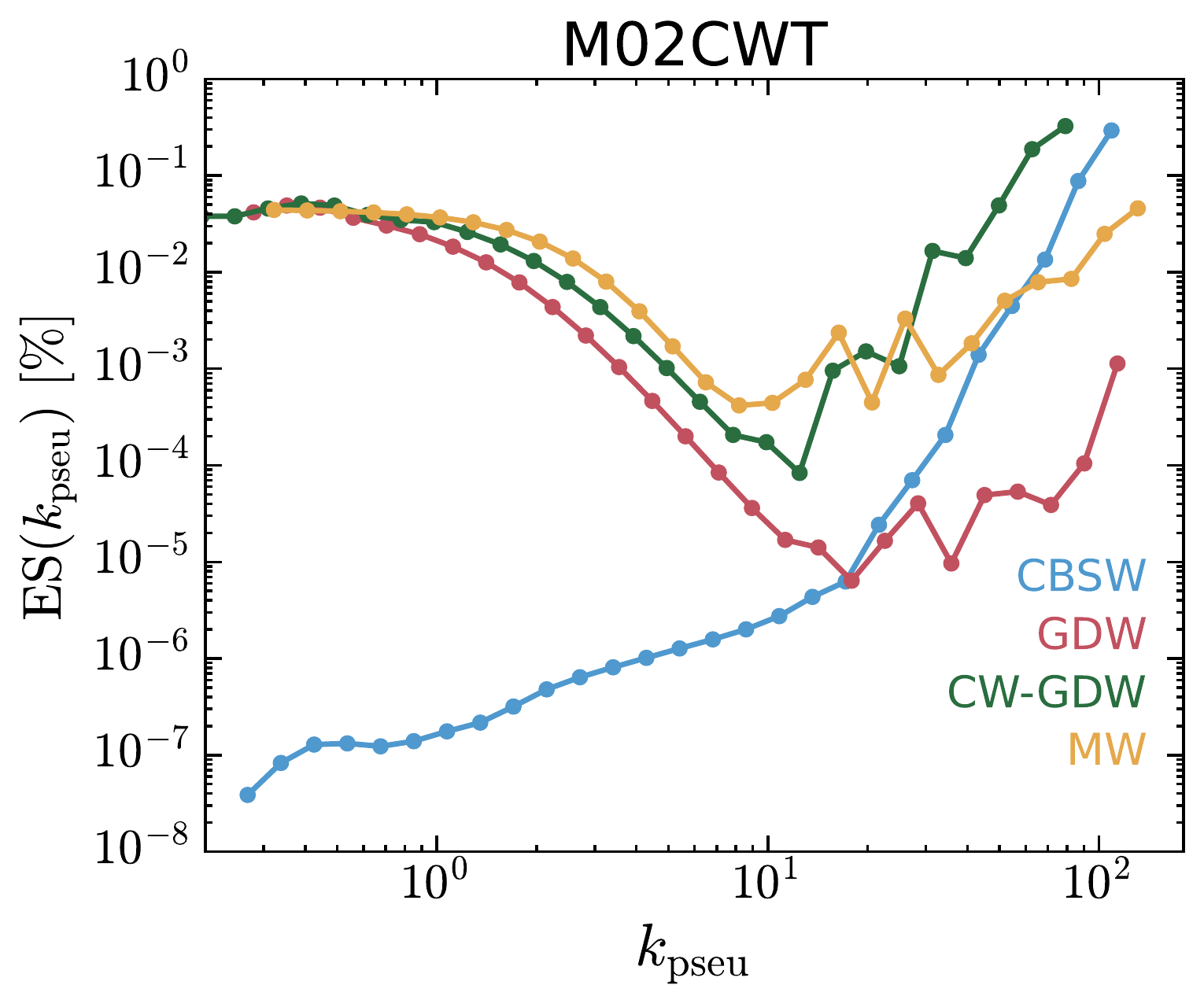}}
	\subfigure{\includegraphics[width=0.44\textwidth]{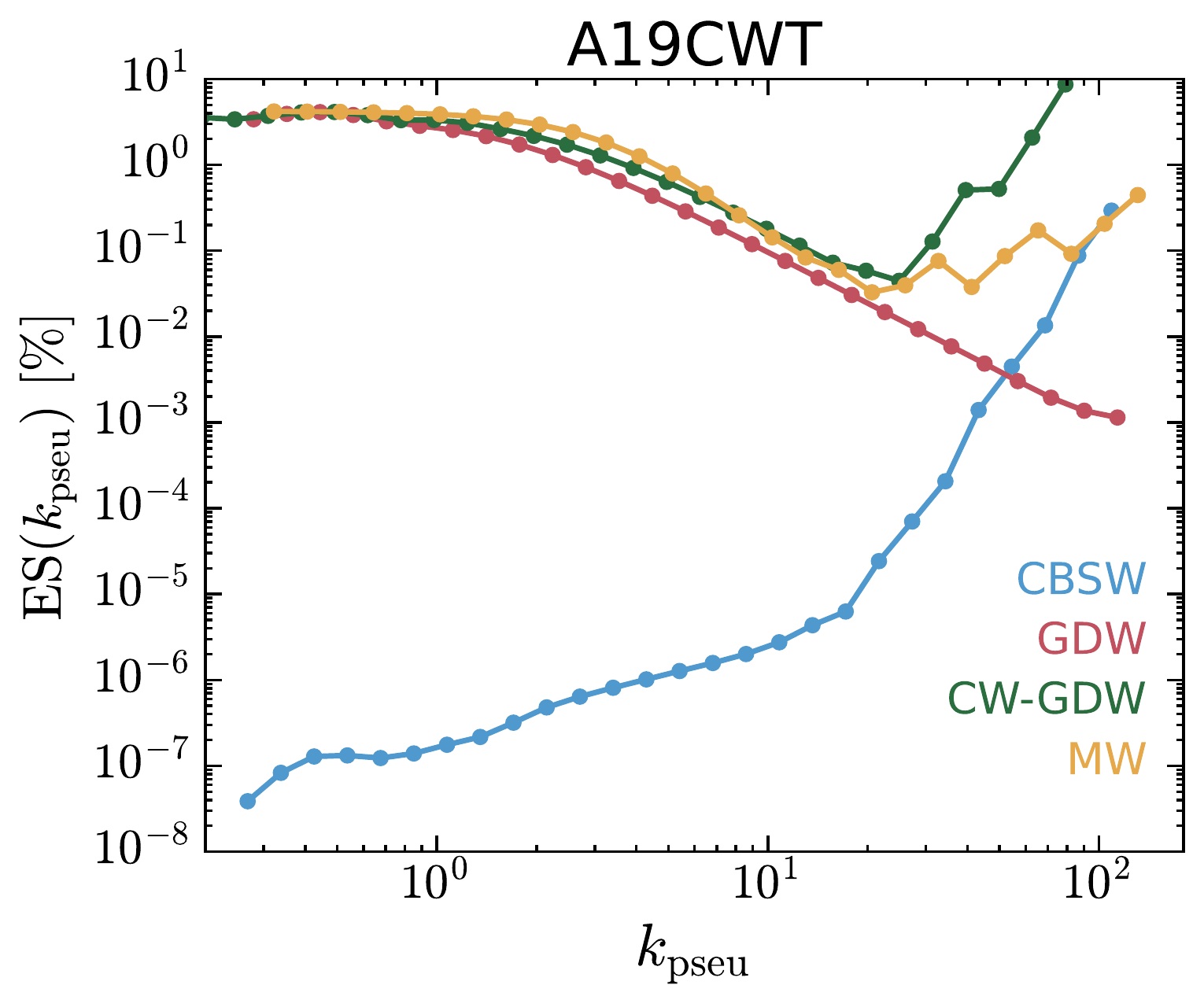}}
	\caption{Same as Fig. \ref{fig:es_spectra_periodic}, but for the results of the non-periodic signal $f_2(x)$.}
	\label{fig:es_spectra_compact}
\end{figure*}

\begin{figure*}
	\centering
	\subfigure{\includegraphics[width=0.45\textwidth]{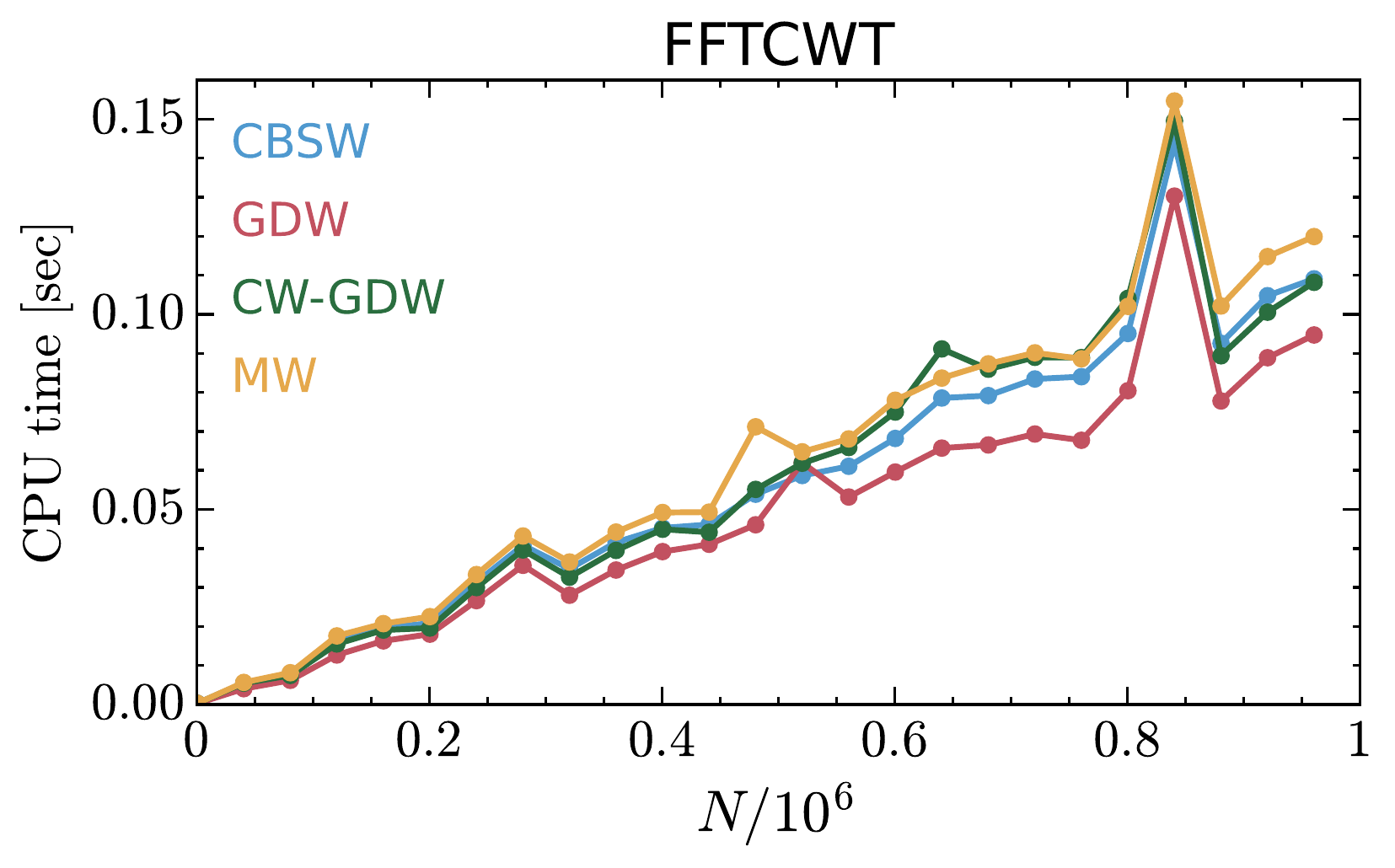}}
	\subfigure{\includegraphics[width=0.45\textwidth]{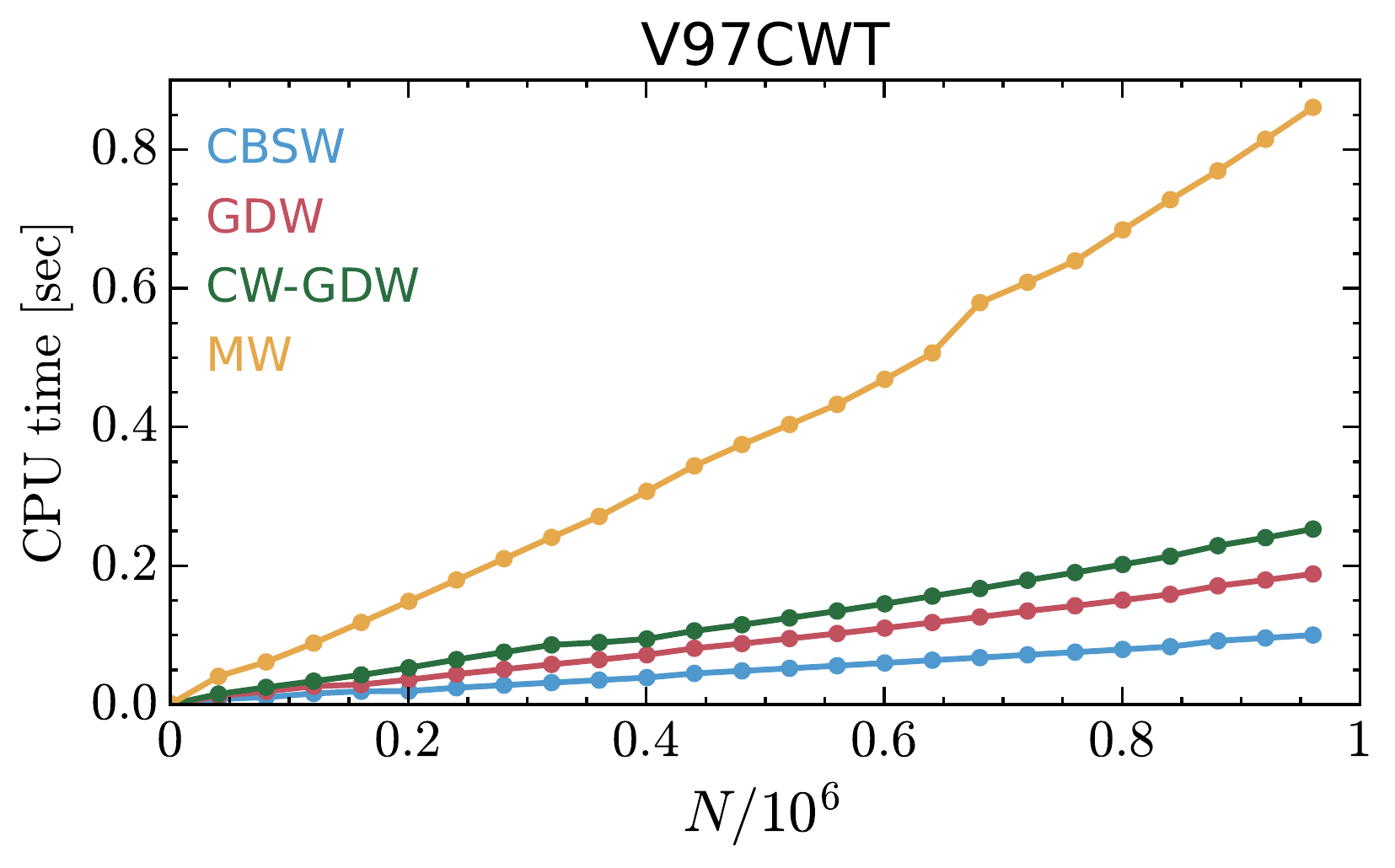}}
	\subfigure{\includegraphics[width=0.45\textwidth]{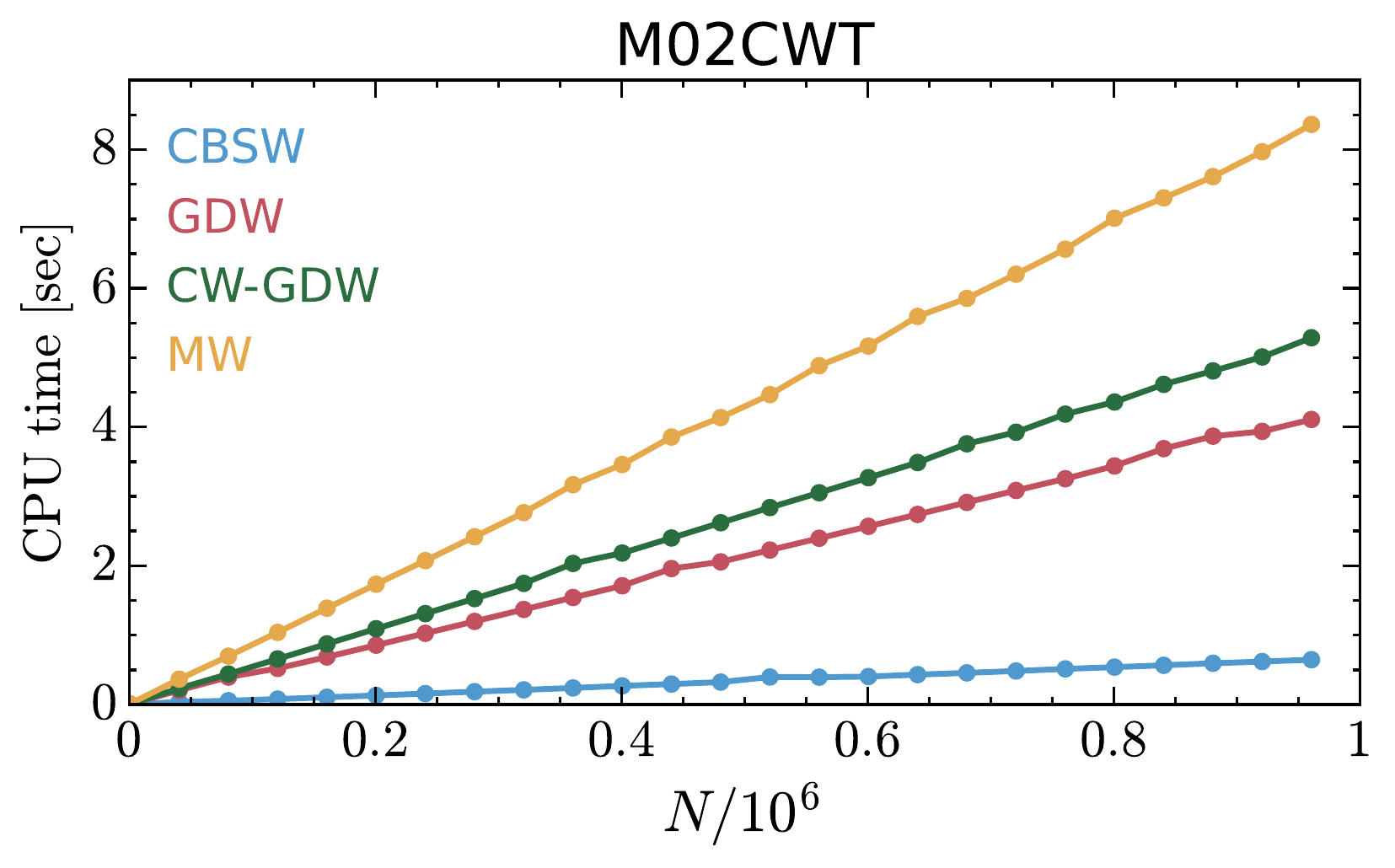}}
	\subfigure{\includegraphics[width=0.45\textwidth]{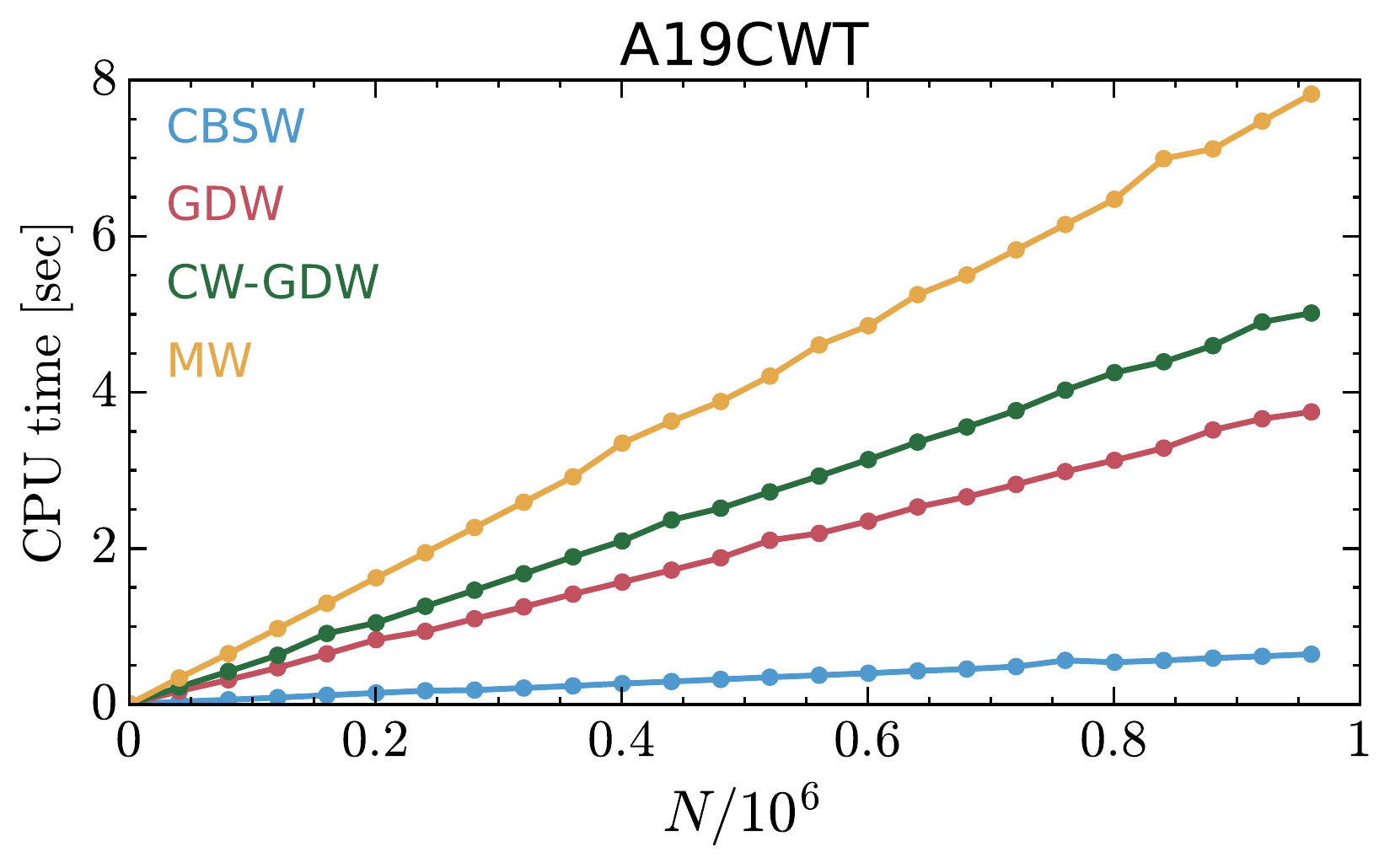}}
	\caption{The CPU time per scale of the different algorithms with different wavelets to compute the numerical CWT of the periodic signal $f_1(x)$.}
	\label{fig:run_time_periodic}
\end{figure*}

\begin{figure*}
	\centering
	\subfigure{\includegraphics[width=0.45\textwidth]{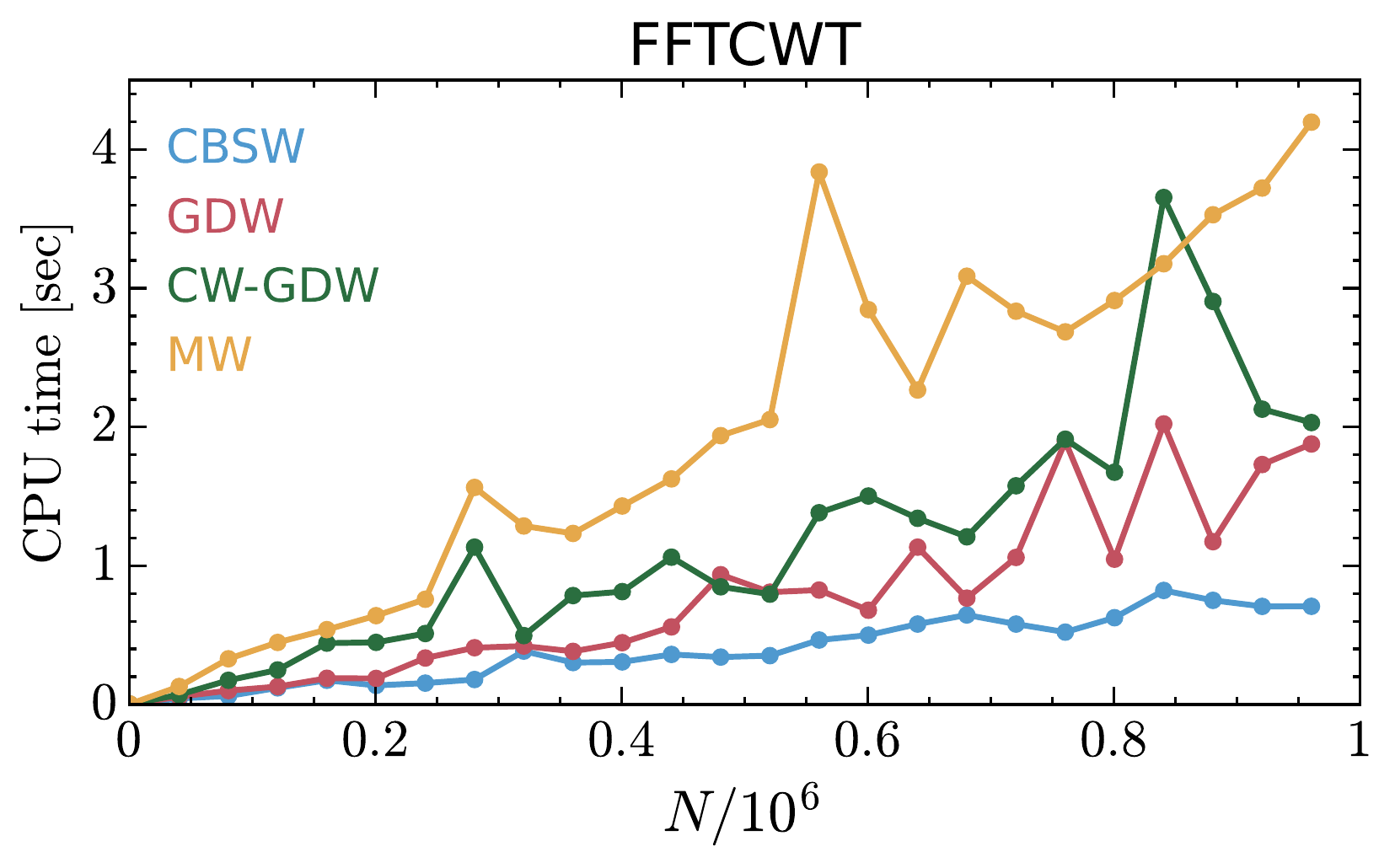}}
	\subfigure{\includegraphics[width=0.45\textwidth]{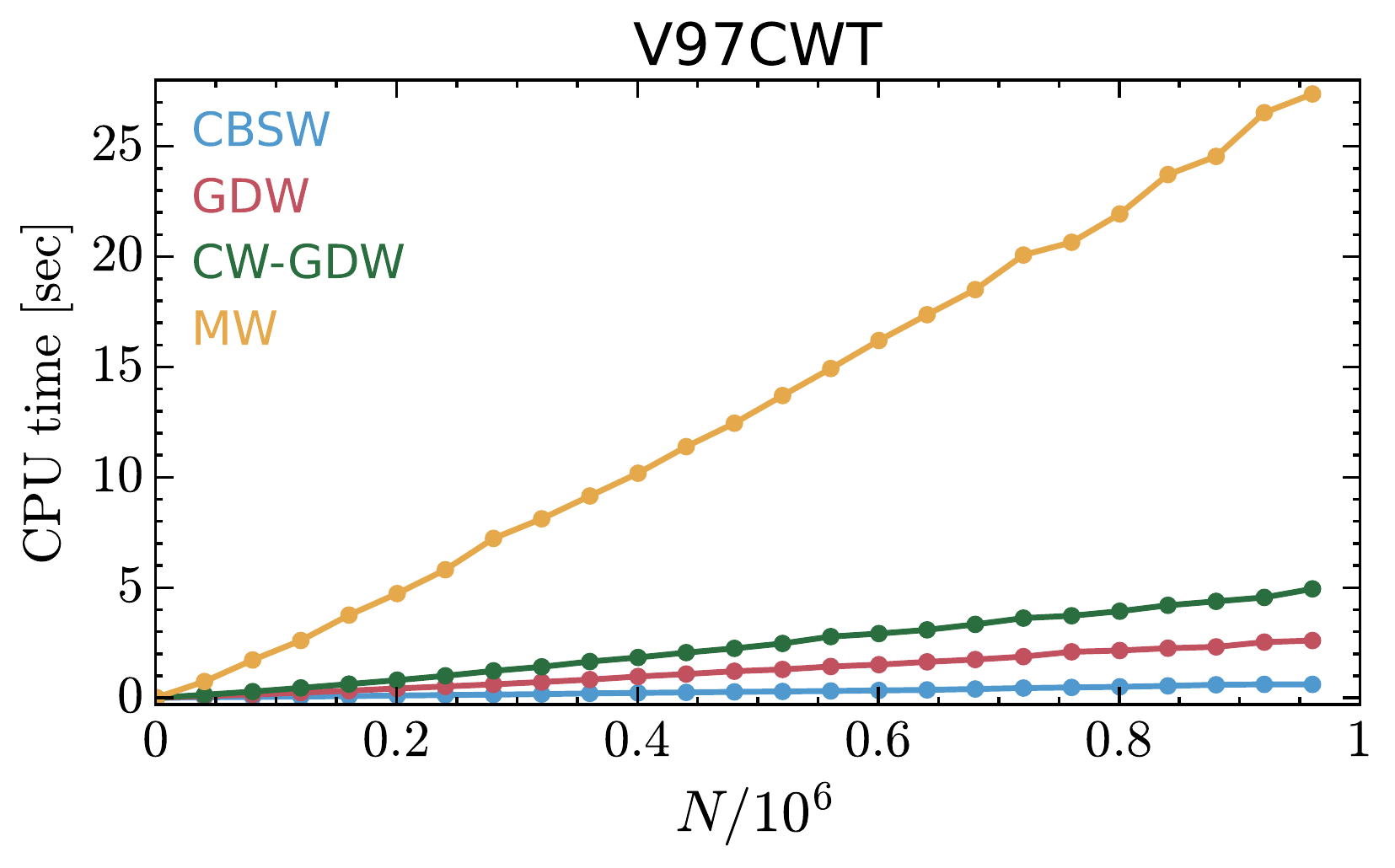}}
	\subfigure{\includegraphics[width=0.45\textwidth]{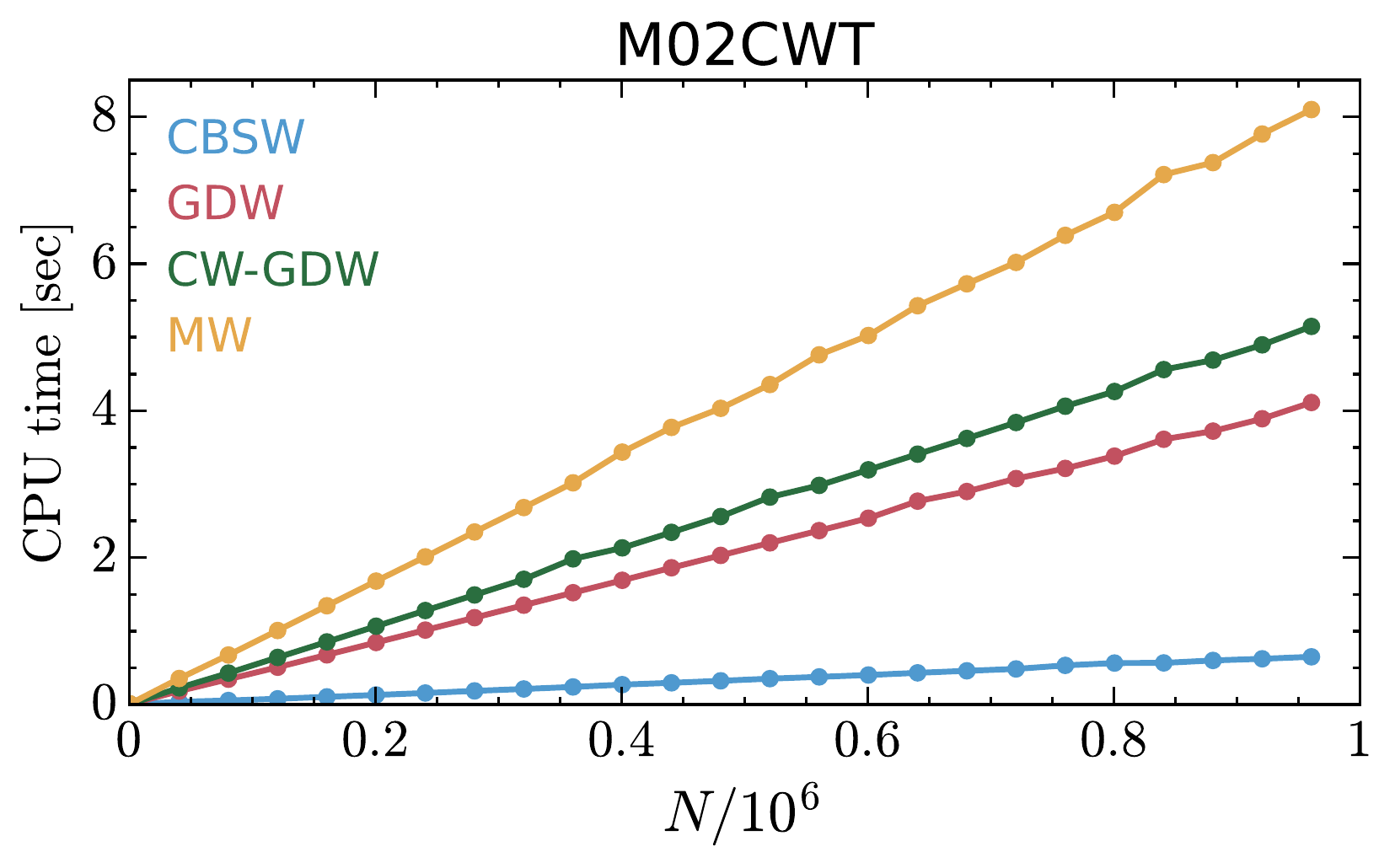}}
	\subfigure{\includegraphics[width=0.45\textwidth]{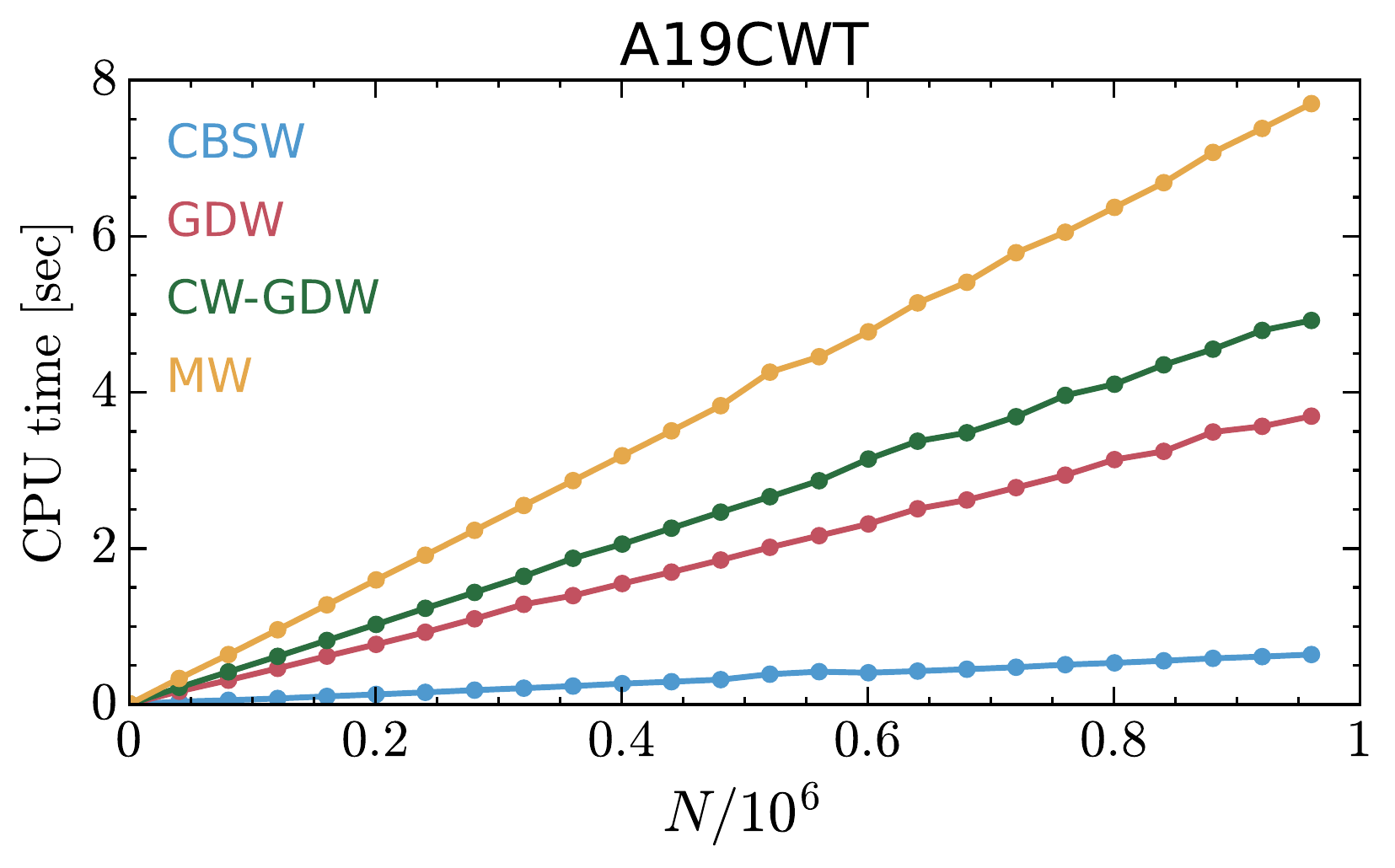}}
	\caption{Same as Fig. \ref{fig:run_time_periodic}, but for the measurements of the non-periodic signal $f_2(x)$.}
	\label{fig:run_time_compact}
\end{figure*}

\section{Performance comparison between algorithms}
\label{sec:performance}

In the 1D case, it is easy to find functions whose CWTs can be calculated analytically by using Equation \eqref{eq:CWT}. Therefore, we can use their analytical results to examine the accuracy of the corresponding numerical outcomes. For instance, we here use the periodic function $f_1(x)$ with period $2\pi$ and the Gaussian function $f_2(x)$ as test signals, which are given below
\begin{align}
f_1(x)&=2\cos(x) + \frac{1}{2}\cos(8x) + \frac{1}{4}\sin(32x),\label{eq:signal_f1}\\
f_2(x)&=e^{-x^2/2}.\label{eq:signal_f2}
\end{align}
The two signals and their analytical CWTs are shown in Fig. \ref{fig:test_signals_cwts}.

For the following numerical tests, we write the double precision codes in Fortran 95 language, and compile them by gfortran 6.3.1 with the \texttt{-O3} flag under the Intel Xeon CPU E5-2678 v3 @ 2.50 GHz processor with 250 GB RAM running Linux (Fedora release 24).

\subsection{Accuracy comparison}
\label{sec:accuracy_comparison}

To check the accuracy of these algorithms, we define the error spectrum as follows
\begin{equation}
\label{eq:error_spectrum}
\mathrm{ES}(w)=\frac{\sum_n|W^\mathrm{n}_f(w,x_n)-W^\mathrm{a}_f(w,x_n)|}{\sum_n |W^\mathrm{a}_f(w,x_n)|}\times 100\%,
\end{equation}
where $W^\mathrm{a}_f(w,x)$ is the analytical CWT, and $W^\mathrm{n}_f(w,x)$ is the numerical CWT.

The computation of the numerical CWT $W^\mathrm{n}_f(w,x)$ requires the sampling of the signal. For the signals $f_1(x)$ and $f_2(x)$, we take
$N=512$ evenly spaced sample points on the intervals $[0,2\pi)$ and $[-6,6)$, respectively, which is sufficient to avoid the aliasing effect. The periodic boundary condition is used for $f_1(x)$, and the zero boundary condition for $f_2(x)$. Since the FFTCWT and V97CWT always assume the signals are periodic, we should pad zeros at both ends of the signal before execute the CWT of $f_2(x)$. Let $N_\mathrm{zeros}$ denote the number of padded zeros at each end, then it can be determined by $\chi$ and $\tilde{w}_\mathrm{min}$:
\begin{align}
\label{eq:num_padding_zeros}
N_\mathrm{zeros}&=\mathrm{Nint}(\chi/\tilde{w}_\mathrm{min}),\nonumber\\
&= \mathrm{Nint}\big(\frac{\chi}{c_w\pi}\big)N.
\end{align}
However, padding zeros takes up more computational resources and reduce the efficiency of the algorithm, which we will see in the next subsection.

In Fig. \ref{fig:es_spectra_periodic}, we show the error spectra of the periodic signal $f_1(x)$. We see that the FFTCWT algorithm yields the highest accuracy, the error of which is less than $10^{-10}\%$. The error of the V97CWT algorithm is between $0.01\%$ and $1\%$. The errors of these two algorithms do not show any significant dependence on the kinds of wavelets. We observe that for the CBSW, the errors of both M02CWT and A19CWT are approximately between $3\times10^{-9}\%$ and $0.003\%$. But for other wavelets, the errors are clearly higher than that for the CBSW, which are ranging from $0.003\%$ to $1\%$ in the case of the M02CWT, and from $0.1\%$ to $10\%$ in the case of the A19CWT. The larger error of the A19CWT may be due to the too coarse approximation in Equation \eqref{eq:3rd_derivative_wavelet}. If $N_\chi$ is larger, this approximation will be more accurate, but the A19CWT will be less efficient.

In Fig. \ref{fig:es_spectra_compact}, we show the error spectra of the non-periodic signal $f_2(x)$. We observe that the error magnitudes of the FFTCWT ($10^{-8}\%-0.1\%$) in handling the non-periodic signal $f_2(x)$ are much higher than that ($10^{-14}\%-10^{-10}\%$) in handling the periodic signal $f_1(x)$, whereas the error magnitudes of the other algorithms do not change much. Even so, the FFTCWT still provides the best accuracy among all algorithms for all wavelets. Only for the CBSW, the accuracy of the M02CWT and A19CWT can rival the accuracy of the FFTCWT.

In summary, the V97CWT, M02CWT, and A19CWT algorithms, which perform CWT calculations in real space, are not as precise as the FFTCWT. The reason for this is mainly that the former threes make approximations to the wavelet function to trade off the efficiency, but yet the latter does not. For the general wavelets, A19CWT yields the largest error, which is due to twice approximations, namely Equations \eqref{eq:piecewise_polys_psi} and \eqref{eq:3rd_derivative_wavelet}, as stated above. However, for the special wavelet CBSW, the M02CWT and A19CWT algorithms provide a quite high accuracy owing to the fact that Equations \eqref{eq:beta3_wavelet_approx_m02}, \eqref{eq:piecewise_polys_psi} and \eqref{eq:3rd_derivative_wavelet} describe the CBSW exactly.

\subsection{Speed comparison}

To check the actual efficiency of the algorithms, we measured the variation of their CPU time with the number of sampling points per scale.

Fig. \ref{fig:run_time_periodic} shows the measurements of the periodic signal $f_1(x)$. We see that the V97CWT, M02CWT and A19CWT algorithms without using the FFT are indeed very fast and they all have the complexity of $\mathcal{O}(N)$ but with different leading constants. However, they do not show a huge speed advantage over the FFTCWT at the sampling number of $N\lesssim 10^{6}$, which can be due to two reasons. On the one hand, the FFT library we used, \texttt{FFTW}, is very well optimized, and is the fastest free library available for computing the FFT. On the other hand, the leading constants of the V97CWT, M02CWT and A19CWT are too large. The V97CWT performances better than the M02CWT and A19CWT, due to its recursive nature. Its CPU time are comparable to that of the FFTCWT for the real wavelets.

Fig. \ref{fig:run_time_compact} shows the measurements of the non-periodic signal $f_2(x)$. It is clearly seen that the FFTCWT and V97CWT consume much more computational time than they do in processing the periodic signal, since we pad many zeros to the signal $f_2(x)$ before executing the CWT. Only for the complex wavelet, MW, the FFTCWT still maintains the speed advantage over the other algorithms. For other wavelets, the FFTCWT has no distinct speed advantage. The CPU time consumed by the M02CWT and A19CWT do not differ by whether the signal is periodic or non-periodic.

\begin{figure*}
	\centerline{\includegraphics[width=0.98\textwidth]{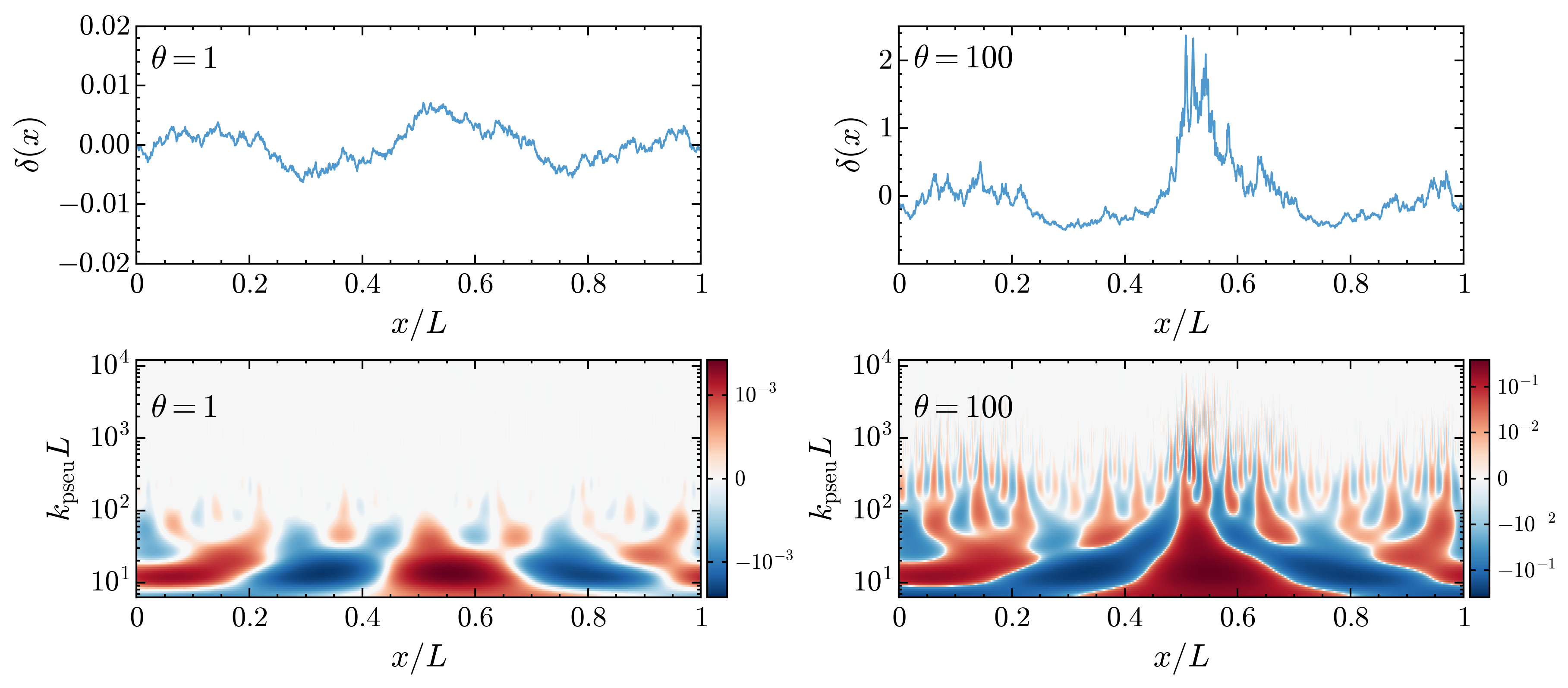}}
	\caption{\textit{Top row}: the initial density field (left panel) and the nonlinear density field at $\theta=100$ (right panel). \textit{Bottom row}: the corresponding CWTs of the density fields, which are computed by using the FFTCWT algorithm with the CW-GDW. In these plots, the coordinates $x$ and scales $k_\mathrm{pseu}$ are made dimensionless by dividing and multiplying the length size $L$ of the density field, respectively.   }
	\label{fig:dens_cwts}
\end{figure*}

\begin{figure}
	\centerline{\includegraphics[width=0.48\textwidth]{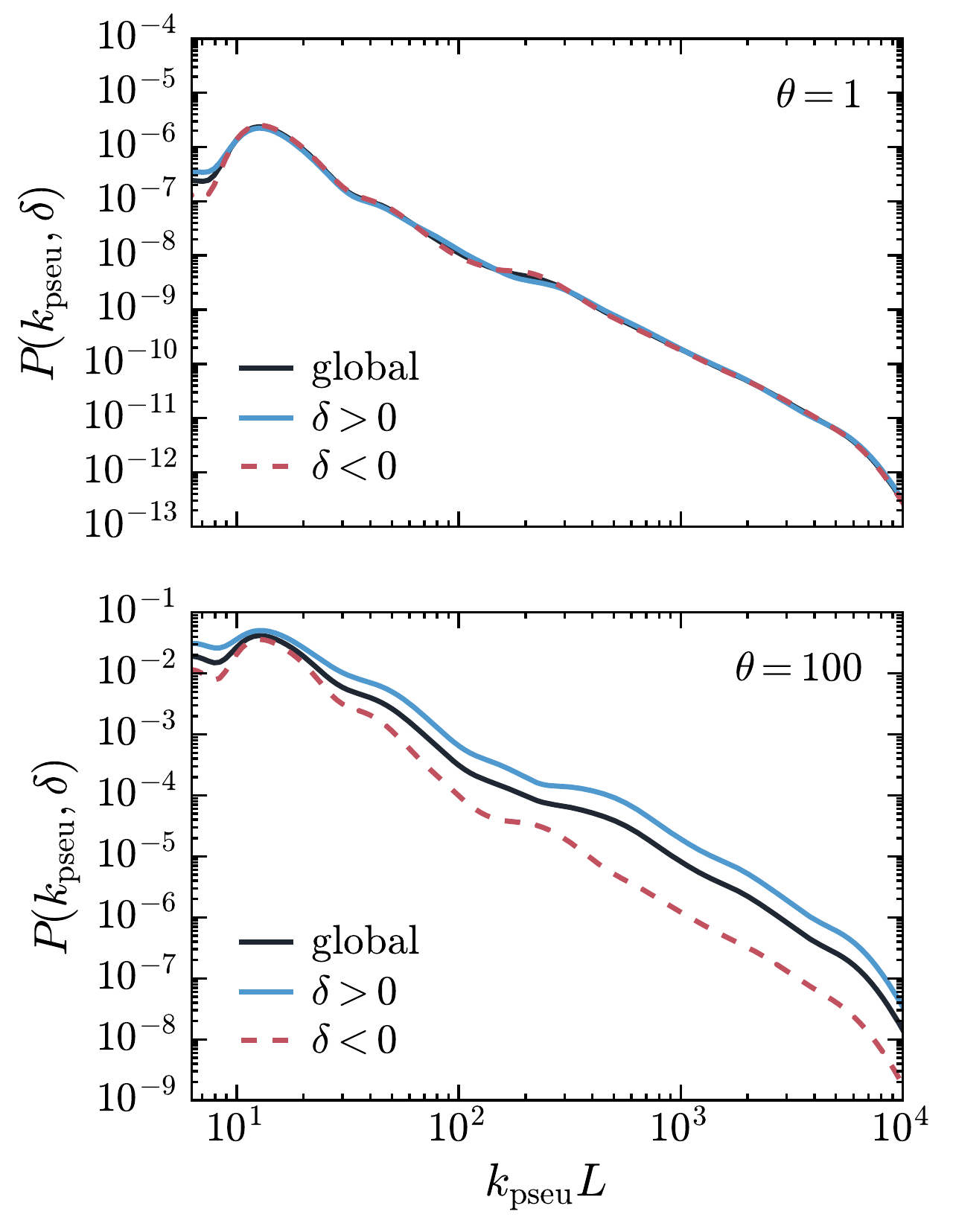}}
	\caption{Comparison of the env-WPS of the initial density field and that of
		the late time density field. \textit{Top panel}: the env-WPS of the initial
		density field with power-law power spectrum $P(k)\propto k^{-2}$. \textit{Bottom panel}: the
		env-WPS of the nonlinear density field at $\theta = 100$. In each panel, the global WPS is denoted by the black line.}
	\label{fig:env_wps}
\end{figure}

\section{Applications in cosmology}
\label{sec:cosmo_apps}

There are many 1D signals in the astrophysics and cosmology, such as the light curves of astronomical sources, the Lyman-$\alpha$ forest, the 21 cm signal, the gravitational waves, and the cosmic fields obtained by solving 1D perturbative equations. The CWT can map them into the 2D time-frequency or space-scale domains, which reveals totally the complex and irregular structures at various positions and scales. Furthermore, we can construct some statistics based on the CWT to characterize the signals more quantitatively, for example, the wavelet power spectrum, the wavelet cross-correlation, the wavelet bicoherence, the wavelet modulus maxima and so on \citep[e.g.][]{Muzy1991, Hudgins1993, vanMilligen1995a, vanMilligen1995b}.

As illustration, we perform the wavelet analysis of the density fields obtained by the 1D Zel'dovich approximation, which provides the exact nonlinear
solution for the perturbative equations of collisionless matter up to the first appearance of orbit-crossing singularities. The nonlinear density field is given by
\begin{equation}
\label{eq:nonlinear_dens}
\delta(x,\theta)+1  = \frac{1}{1-\theta \delta_0(x)},
\end{equation}
where $\theta$ is the growth factor and used as the time variable, and $\delta_0(x)=\delta(x,\theta=1)$ is the initial Gaussian density field satisfying periodic boundary conditions, which is generated by the power-law spectrum $P(k)=Ak^{-2}$ with $A=2.5\times10^{-6}$. For more details about the 1D Zel'dovich approximation, we refer the reader to \citet{Wang2022a}.

From the results in Section \ref{sec:accuracy_comparison}, it is clear that the FFTCWT algorithm is optimal for the periodic signal. In addition, as can be seen from Fig. \ref{fig:wavelets}, the CW-GDW achieves a better balance between spatial resolution and scale resolution compared to other wavelets. Therefore, we compute the CWTs of density fields by the FFTCWT algorithm with the CW-GDW, the results of which are illustrated in Fig. \ref{fig:dens_cwts}. By visual inspection, the CWT of the initial density field is dominated by large-scale components with a relatively random spatial distribution. As a consequence of the nonlinear gravitational effect, the CWT of the density field at $\theta = 100$ shows a non-random structure with many small-scale components, which do not exist at the initial time.

In our previous work \citep{Wang2022b}, we proposed the environment-dependent wavelet power spectrum (env-WPS) to measure the dependence of matter clustering on both the scale and environment, which is given by
\begin{equation}
\label{eq:env_wps}
P(w,\delta')=\langle|W_\delta(w,x)|^2 \rangle_{\delta(x)=\delta'},
\end{equation}
where $W_\delta(w,x)$ is the CWT of $\delta(x)$, and ``$\langle\ldots\rangle_{\delta(x)=\delta'}$" denotes the statistical average of the
wavelet coefficients at each scale with the same local density, i.e. $\delta(x)=\delta'$. If we average over all the possible densities, then the env-WPS will degenerate to the global WPS as bellow
\begin{equation}
\label{eq:global_wps}
P(w)=\langle|W_\delta(w,x)|^2 \rangle_{\mathrm{all}\ \delta'}.
\end{equation}
Thus the relation between the global WPS $P(w)$ and the env-WPS $P(w,\delta')$ is
\begin{equation}
\label{eq:global_wps_env_wps}
P(w)=\sum_{\delta'}f_{\delta'}P(w,\delta'),
\end{equation}
where $f_{\delta'}=N_{\delta'}/N$ is the fraction of the env-WPS relative to the global WPS, and $N_{\delta'}$ is the number of grids at $\delta(x)=\delta'$. In fact, the env-WPS can be generalized to other kinds of signals, just by replacing $\delta'$ with the corresponding attribute.

For simplicity, we here split the densities into: (a) $\delta>0$, i.e. the overdense environments and (b) $\delta<0$, i.e. the underdense environments, and then compute the env-WPSs, the results of which are shown in Fig. \ref{fig:env_wps}. For the initial density field, we can see that its env-WPSs have the same amplitudes with the global WPS. However, for the fully evolved density field at $\theta=100$, the env-WPSs exhibit an obvious environment dependence. Specifically, the env-WPS with $\delta>0$ is larger than the global WPS, while that with $\delta<0$ is less than the global WPS.

As can be seen, the env-WPS provides more information about matter clustering than the traditional two-point statistics, e.g. FT-based power spectrum, which completely lost the characteristics of the matter spatial distribution.\footnote{\citet{Wang2022b} makes it more explicitly.}

\section{Summary and Conclusions}
\label{sec:conclusions}

In this paper, we review the fast algorithms for the CWT, including the FFTCWT with complexity of $\mathcal{O}(N\log_2N)$ per scale, and other three algorithms with complexity of $\mathcal{O}(N)$ per scale, i.e. the V97CWT proposed by \citet{Vrhel1997}, the M02CWT proposed by \citet{Munoz2002}, and the A19CWT proposed by \citet{Arizumi2019}.

By the convolution theorem, the FFTCWT converts convolution calculations in the real domain into multiplications in the Fourier domain, and then returns the final result by the inverse FT (see Section \ref{sec:fftcwt_algo} and Fig. \ref{fig:fftCWT}). The V97CWT use the daughter wavelets with scales of $\tilde{w}_\mathrm{max}/2^{j/N_\mathrm{subs}}$ and approximate them with two scaling functions, i.e. the zero order and the cubic B-splines. Using the two-scale relation of the B-splines, then the CWT can be calculated recursively (see Section \ref{sec:v97cwt_algo} and Fig. \ref{fig:Vrhel97}). The M02CWT approximate the daughter wavelet with the rescaled cubic B-spline, and interpolating the discrete input signal with the cubic B-spline. Therefore, the large wavelet convolution kernel is translated into smaller B-spline kernel (see Section \ref{sec:m02cwt_algo} and Fig. \ref{fig:Munoz02}). The A19CWT achieves the same purpose as the M02CWT by approximating the mother wavelet as cubic piecewise polynomials and applying integration by parts (see Section \ref{sec:a19cwt_algo} and Fig. \ref{fig:Arizumi19}). In fact, the precision of algorithms originally mentioned in \citet{Munoz2002} and \citet{Arizumi2019} is terrible on small scales, and we remedy this issue in our M02CWT and A19CWT algorithms (see Appendix \ref{sec:m02cwt_accuracy}).

We compare the accuracy and speed between these fast CWT algorithms in Figs. \ref{fig:es_spectra_periodic}-\ref{fig:run_time_compact} by using two specific signals. Our main findings are summarized as follows:

\begin{enumerate}[leftmargin=2\parindent]
	\item Even though for the non-periodic signal with zero boundaries, the accuracy of the FFTCWT is much lower compared to that for the periodic signal, it is still more accurate than other algorithms.
	\item When the $\mathcal{O}(N)$ algorithms process the non-periodic signal with zero boundaries, the overall magnitudes of their errors do not grow larger compared to when they process the periodic signal. Hence the accuracy of them is robust to different types of signals. The M02CWT achieves the best accuracy among them.
	\item For the GDW, CW-GDW and MW, A19CWT is the least accurate algorithm. But for the CBSW, the A19CWT is just as accurate as the M02CWT, which is because both the cubic B-spline and piecewise polynomials represent the CBSW exactly.
	\item At the sampling number we consider, i.e. $N\lesssim 10^{6}$, the algorithms with the complexity of $\mathcal{O}(N)$ per scale do not exhibit an overall speed advantage over the FFTCWT. Only the V97CWT with real wavelets shows a speed comparable to it.
	\item For the non-periodic signal with zero boundaries, the FFTCWT and V97CWT are less efficient due to padding zeros to the signal. However, the efficiency of the M02CWT and A19CWT  is not affected by the type of signals.
\end{enumerate}
Therefore, the FFTCWT and V97CWT are suitable for the periodic signals. In particular, the V97CWT using real wavelets will perform better than using complex wavelets, e.g. the MW. The M02CWT is suitable for the non-periodic signals with zero boundary condition. We do not refer to the A19CWT algorithm because it is not accurate enough.

As a demonstration of the usage of the CWT, we then apply the FFTCWT to perform wavelet analysis of the 1D density fields. Acting like a ``mathematical microscope", the CWT allows us to zoom in on complex structures of the density fields at various scales and locations (see Fig. \ref{fig:dens_cwts}). We also introduce the wavelet-based statistic, env-WPS, which is a bivariate function of the local density environment and the scale. As shown in Fig. \ref{fig:env_wps}, the env-WPS tells us that for the initial field, there is no any environment dependence of matter clustering on all scales. However, for the late time field, the matter clustering is dominated by the matter in overdense environments. Clearly, the env-WPS contains more information about the matter clustering than the usual two-point statistics. The env-WPS can also be generalized to analyze other signals by replacing the local density environment by other attribute.

To analyze the multi-dimensional data, such as the 2D gravitational lensing maps and the 3D cosmic fields, the next natural step is to extend the 1D CWT algorithms to multi-dimensions. It is ease to develop 2D and 3D FFT-based CWT algorithms, since there are publicly available FFT libraries to use. However, multi-dimensional extensions of the rest 1D algorithms are not straightforward. In the future, we will plan to develop the fast multi-dimensional CWT algorithms without the use of FFT.

\section*{Acknowledgments}

Y.W. especially thanks Dr. Vrhel and Mr. Arts for helpful discussions. P.H. acknowledges the support by the National Science Foundation of China (No. 12047569, 12147217), and by the Natural Science Foundation of Jilin Province, China (No. 20180101228JC).

\section*{Data Availability}
The double precision Fortran 95 codes of the FFTCWT, V97CWT, M02CWT, and A19CWT algorithms are released in \url{https://github.com/WangYun1995/FortranCWT}. The corresponding Python wrappers are available in \url{https://github.com/WangYun1995/pyFortranCWT}.

\appendix

\section{Derivation of the simple inversion formula for the CWT}
\label{sec:ICWT}

In our previous works \citep{Wang2021,Wang2022a}, we demonstrate that there exists a single integral inverse formula for the real-valued wavelet derived by smoothing window function which is shown below
\begin{equation}
\label{eq:ICWT_smooth}
f(x) = f(w\rightarrow0,x)+\int_0^{+\infty}\frac{W_f(w,x)}{\sqrt{w}}\mathrm{d}w,
\end{equation}
where $f(w\rightarrow0,x)=\Lim{w\rightarrow 0}\int f(u)S(w,x-u)\mathrm{d}u$, $S(w,x)=wS(wx)$ is a smoothing function with scale $w$, and $W_f(w,x)$ is the CWT of $f(x)$ based on the wavelet $\psi(w,x)=\sqrt{w}\partial S(w,x)/\partial w$.

In fact, we can generalize Equation \eqref{eq:ICWT_smooth} to hold for more general real wavelets. According to the convolution theorem, the CWT $W_f(w,x)$ can be expressed as
\begin{displaymath}
W_f(w,x) = \frac{1}{2\pi}\int_{-\infty}^{+\infty}\hat f(k)\frac{1}{\sqrt{w}}\hat\psi\left(\frac{k}{w}\right)e^{-ikx}\mathrm{d}k,
\end{displaymath}
where $\hat f(k)$ and $\hat\psi(k)$ are Fourier transforms of $f(x)$ and $\psi(x)$, respectively. Divide the L.H.S. and R.H.S. of the above equation by $\sqrt{w}$ and integrate over $w$, we obtain
\begin{align}
\int\limits_0^{+\infty}\frac{W_f(w,x)}{\sqrt{w}}\mathrm{d}w &=  \frac{1}{2\pi}\int\limits_{-\infty}^{+\infty}\left(\int\limits_0^{+\infty}\frac{1}{w}\hat\psi(\frac{k}{w})\mathrm{d}w \right)\hat f(k)e^{-ikx}\mathrm{d}k.
\end{align}
Let's observe the value of $\int_0^{+\infty}\frac{1}{w}\hat\psi(\frac{k}{w})\mathrm{d}w$:
\begin{enumerate}[leftmargin=2\parindent]
	\item If $k=0$, it follows from the oscillatory nature of wavelets $\hat\psi(0)=\int_{-\infty}^{+\infty}\psi(x)\mathrm{d}x=0$ that $\int_0^{+\infty}\frac{1}{w}\hat\psi(\frac{k}{w})\mathrm{d}w=0$. Hence, the zero frequency component of $\hat f(k)$ is subtracted by CWT.
	\item If $k>0$ and let $u=k/w$, we have $\int_0^{+\infty}\frac{1}{w}\hat\psi(\frac{k}{w})\mathrm{d}w=\int_0^{+\infty}\frac{1}{u}\hat\psi(u)\mathrm{d}u$.
	\item If $k<0$ and let $u=-k/w$, we have $\int_0^{+\infty}\frac{1}{w}\hat\psi(\frac{k}{w})\mathrm{d}w=\int_0^{+\infty}\frac{1}{u}\hat\psi(-u)\mathrm{d}u$.
\end{enumerate}

It is clear from the above that if the real-valued wavelets satisfy
\begin{equation}
\label{eq:even_cond}
\psi(x) = \psi(-x)
\end{equation}
which is equivalent to $\hat\psi(k)=\hat\psi(-k)$, and
\begin{equation}
\label{eq:new_admis}
0<\left|\mathcal{K}_\psi\equiv\int_0^{+\infty}\frac{1}{k}\hat\psi(k)\mathrm{d}k\right|<\infty,
\end{equation}
then we get
\begin{align*}
\frac{1}{\mathcal{K}_\psi}\int_0^{+\infty}\frac{W_f(w,x)}{\sqrt{w}}\mathrm{d}w &= \frac{1}{2\pi}\int\limits_{k\neq0}\hat f(k)e^{-ikx}\mathrm{d}k\\
&=\frac{1}{2\pi}\int_{-\infty}^{+\infty}\hat f(k)e^{-ikx}\mathrm{d}k \\
&\quad -\lim_{\delta k \rightarrow 0} \frac{1}{2\pi}\int_{{-}\delta k/2}^{{+}\delta k/2}\hat f(k)e^{-ikx}\mathrm{d}k \\
&= f(x)-\lim_{\delta k \rightarrow 0}\frac{\delta k}{2\pi}\hat f(0)\\
&= f(x)-\lim_{L \rightarrow \infty}\frac{1}{L}\int_{-L/2}^{L/2}f(x)\mathrm{d}x.
\end{align*}
Finally, we arrive at the simple inversion formula for the CWT, which is
\begin{equation}
f(x)=\bar{f}+\frac{1}{\mathcal{K}_\psi}\int_0^{+\infty}\frac{W_f(w,x)}{\sqrt{w}}\mathrm{d}w,
\end{equation}
where $\bar{f}=\Lim{L \rightarrow \infty}\frac{1}{L}\int_{-L/2}^{L/2}f(x)\mathrm{d}x$ denotes the average of $f(x)$ over all space. For example, in the case of periodic functions, $\bar{f}$ is equal to the average of the function over a period. In the case of compactly supported functions, $\bar{f}$ is equal to zero.

\section{B-spline functions}
\label{sec:bsplines}
The B-spline function of degree zero $\beta^0(x)$ is defined as
\begin{equation}
\label{eq:beta0}
\beta^0(x)=
\begin{cases}
1, & 1/2\leq x\leq1/2,\\
0, & \mathrm{otherwise},
\end{cases}
\end{equation}
and the B-spline $\beta^n(x)$ of degree $n$ is constructed from the $n$ times convolution of $\beta^0(x)$:
\begin{equation}
\label{eq:betan}
\beta^n(x) = \big(\underbrace{\beta^0*\beta^0*\ldots *\beta^0}_{n\ \text{times}}\big)(x).
\end{equation}
Obviously, the B-spline $\beta^{n_1+n_2}$ can be calculated by convolving the B-splines $\beta^{n_1}$ and $\beta^{n_2}$ as follows
\begin{equation}
\label{eq:two_betas}
\beta^{n_1+n_2}(x)=(\beta^{n_1}*\beta^{n_2})(x).
\end{equation}

B-splines have many useful properties, which are listed below
\begin{enumerate}[leftmargin=2\parindent]
	\item They are compactly supported functions with support interval $[-(n+1)/2,(n+1)/2]$ \citep{Briand2018}.
	\item They satisfy a two-scale relation \citep{Vrhel1997}, which is
	\begin{equation}
	\label{eq:two_scale}
	\beta^n(x/2)=\sum_m h(m)\beta^n(x-m),
	\end{equation}
	where the coefficients $h(m)$ are given by
	\begin{align}
	\label{eq:h_filter}
	h(m)=
	\begin{cases}
	\frac{1}{2^n}\binom{n+1}{m+(n+1)/2}, & |m|\leq (n+1)/2,\\
	0, & \mathrm{otherwise}.
	\end{cases}
	\end{align}
	\item The rescaled B-spline of degree $n$ is \citep{Munoz2002}
	\begin{align}
	\label{eq:rescaled_beta}
	w\beta^n(wx) &= \nonumber\\
	& w^{n+1}\left(\Delta_w^{n+1}*D^{-(n+1)}\delta^D\left(\cdot+\frac{n+1}{2w}\right)\right)(x),
	\end{align}
	where $\delta^D(x)$ is the Dirac delta function, $D^{-1}$ is the antiderivative (or integral) operator defined as
	\begin{equation}
	\label{eq:integral_operator}
	D^{-1}f(x) = \int_{-\infty}^{x}f(u)\mathrm{d}u,
	\end{equation}
	$\Delta^{n+1}_w$ is the rescaled finite-difference operator defined as
	\begin{align}
	\label{eq:rescaled_finite_diff}
	(\Delta^{n+1}_w*f)(x) &= \sum_{m=0}^{n+1}a(m)f(x-m/w)\nonumber\\
	&= \sum_{m=0}^{n+1}(-1)^m\binom{n+1}{m}f(x-m/w),
	\end{align}
	and $\Delta^{-1}$ is the inverse finite-diference operator defined as
	\begin{equation}
	\label{eq:inverse_finite_diff}
	(\Delta^{-1}*f)(x) = \sum_{m\leq x}f(x-m).
	\end{equation}
	For the discrete signal $f(n)$, its inverse finite-difference $s(n)=(\Delta^{-1}*f)(n)$ can be implemented recursively by
	\begin{equation}
	\label{eq:cumsum_recursively}
	s(n) = s(n-1)+f(n).
	\end{equation}
	\item The $n_1$-th antiderivative of the B-spline of degree $n_2$ is \citep{Munoz2002}
	\begin{equation}
	\label{eq:antiderivative_beta}
	D^{-(n_1)}\beta^{n_2}(x)=\left(\Delta^{-n_1}*\beta^{n_1+n_2}\left(\cdot-\frac{n_1}{2}\right)\right)(x).
	\end{equation}
\end{enumerate}

\section{Implementation of the IIR filter}
\label{sec:iir_filter}

The computation of Equations \eqref{eq:iir_filter_0}, \eqref{eq:iir_filter_i},  and \eqref{eq:interp_coeffs} is essentially to perform IIR filtering on the signal:
\begin{equation}
\label{eq:iir_filter_append}
f_\mathrm{out}(n) = \left(f_\mathrm{in}*[(\beta^n)^{-1}]_{\uparrow m}\right)(n),
\end{equation}
where $f_\mathrm{in}(n)$ is the input discrete signal, $f_\mathrm{out}(n)$ is the filtered signal, $m=2^i$ for Equation \eqref{eq:iir_filter_i}, and $m=1$ for Equations \eqref{eq:iir_filter_0} and \eqref{eq:interp_coeffs}. By performing the $z$-transform \footnote{Please refer \citet{Proakis2007} for the details of the $z$-transform.} on the Equation \eqref{eq:iir_filter_append}, we have
\begin{equation}
\label{eq:iir_z_trans}
\mathcal{F}_\mathrm{out}(z) = \mathcal{B}_{n_0}(z)\mathcal{F}_\mathrm{in}(z),
\end{equation}
where $\mathcal{F}_\mathrm{out}(z)$, $\mathcal{F}_\mathrm{in}(z)$ and $\mathcal{B}_{n_0}(z)$ are the $z$-transforms of $f_\mathrm{out}(n)$, $f_\mathrm{in}(n)$ and $[(\beta^n)^{-1}]_{\uparrow m}(n)$, respectively. According to \citet{Vrhel1997}, the formula of $\mathcal{B}_{n_0}(z)$ is
\begin{equation}
\mathcal{B}_{n_0}(z) = d_0\prod_{j=1}^{n_0}\mathcal{B}(z;z_j),
\end{equation}
in which $\mathcal{B}(z;z_j)$ is defined as
\begin{equation}
\label{eq:B_zj}
\mathcal{B}(z;z_j) = \frac{1}{(1-z_jz^{-m})}\frac{-z_j}{(1-z_jz^m)},
\end{equation}
and $n_0$ is
\begin{equation}
n_0=\mathrm{Floor}(n/2).
\end{equation}
Values of the constant coefficients $d_0$ and $z_j$ are given in \citet{Vrhel1997}.

Therefore, Equation \eqref{eq:iir_z_trans} can be expressed as follows
\begin{align}
\mathcal{F}_0(z) &= \mathcal{F}_\mathrm{in}(z), \label{eq:F0}\\
\mathcal{F}_{j}(z) &= \mathcal{B}(z;z_{j})\mathcal{F}_{j-1}(z),\quad \mathrm{for}\ 1\leq j\leq n_0,\label{eq:iir_z_trans_j}\\
\mathcal{F}_\mathrm{out}(z) &= d_0\mathcal{F}_{n_0}(z)\label{eq:Fn}.
\end{align}
Combining Equation \eqref{eq:B_zj} and \eqref{eq:F0}-\eqref{eq:Fn}, we obtain the following recursive filter equations:
\begin{align}
f_\mathrm{tem}(n) &{=} f_{j-1}(n) {+} z_{j}f_\mathrm{tem}(n{-}m),\ (n{=}m,\ldots,N{-}1)\label{eq:iir_filter_real_a}\\
f_{j}(n) &{=} z_{j}\big(f_{j}(n{+}m){-}f_\mathrm{tem}(n)\big), \ (n{=}N{-}1{-}m,\ldots,0)\label{eq:iir_filter_real_b}
\end{align}
for the input $f_0(n) = f_\mathrm{in}(n)$. Then the output is $f_\mathrm{out}(n)= d_0f_{n_0}(n)$.

To calculate $f_{j}$ recursively, we need to know $f_\mathrm{tem}(n)$ for $n = 0,\ldots,m{-}1$, and $f_{j}(n)$ for $n=N{-}1,\ldots,N{-}m$. By assuming that $f_{j{-}1}$ is periodic over $N$ samples, the initial values can be calculated by
\begin{align}
f_\mathrm{tem}(n) &{=} {\sum_{l=0}^{N_l}}z_{j}^lf_{j-1}[\mathrm{Mod}(n{-}lm,N)],\ (n {=} 0,\ldots,m{-}1), \label{eq:periodic_initial_a}\\
f_{j}(n) &{=} {-}{\sum_{l=0}^{N_l-1}}z_{j}^{l{+}1}f_\mathrm{tem}[\mathrm{Mod}(n{+}lm,N)],\ (n{=}N{-}1,\ldots,N{-}m),\label{eq:periodic_initial_b}
\end{align}
where $N_l=\ln\epsilon/\ln|z_{j}|$, $\epsilon=10^{-16}$ is the prespecified level
of precision, and $\mathrm{Mod}(a,b)$ returns the remainder of the division of $a$ by $b$.

If $n_0=1$ and $m=1$, which is the case for the M02CWT and A19CWT algorithms, then it is also convenient to assume that the signal is zero outside the sampled range. Thus the initial values are
\begin{align}
f_\mathrm{tem}(0) &= f_\mathrm{in}(0),\label{eq:zero_initial_a}\\
f_1(N-1) &= -f_\mathrm{tem}(N-1){\sum_{l=1}^{(N_l{+}1)/2}}z_1^{2l-1}.\label{eq:zero_initial_b}
\end{align}
To calculate the convolution between $f_1$ and $\beta^7$ (e.g. Equations \eqref{eq:m02_cwt} and \eqref{eq:a19cwt_04}), we also need to know
\begin{align}
f_1(n) &= z_1^{-n}f_1(0), \ (n=-6,\ldots,-1)\label{eq:zero_initial_c}\\
f_1(N+n) &= -f_\mathrm{tem}(N-1){\sum_{l=1}^{(N_l{-n})/2}}z_1^{2l+n}, \ (n=0,1)\label{eq:zero_initial_d}.
\end{align}

\section{Accuracy tests of the M02CWT algorithm }
\label{sec:m02cwt_accuracy}

\begin{figure}
	\centerline{\includegraphics[width=0.5\textwidth]{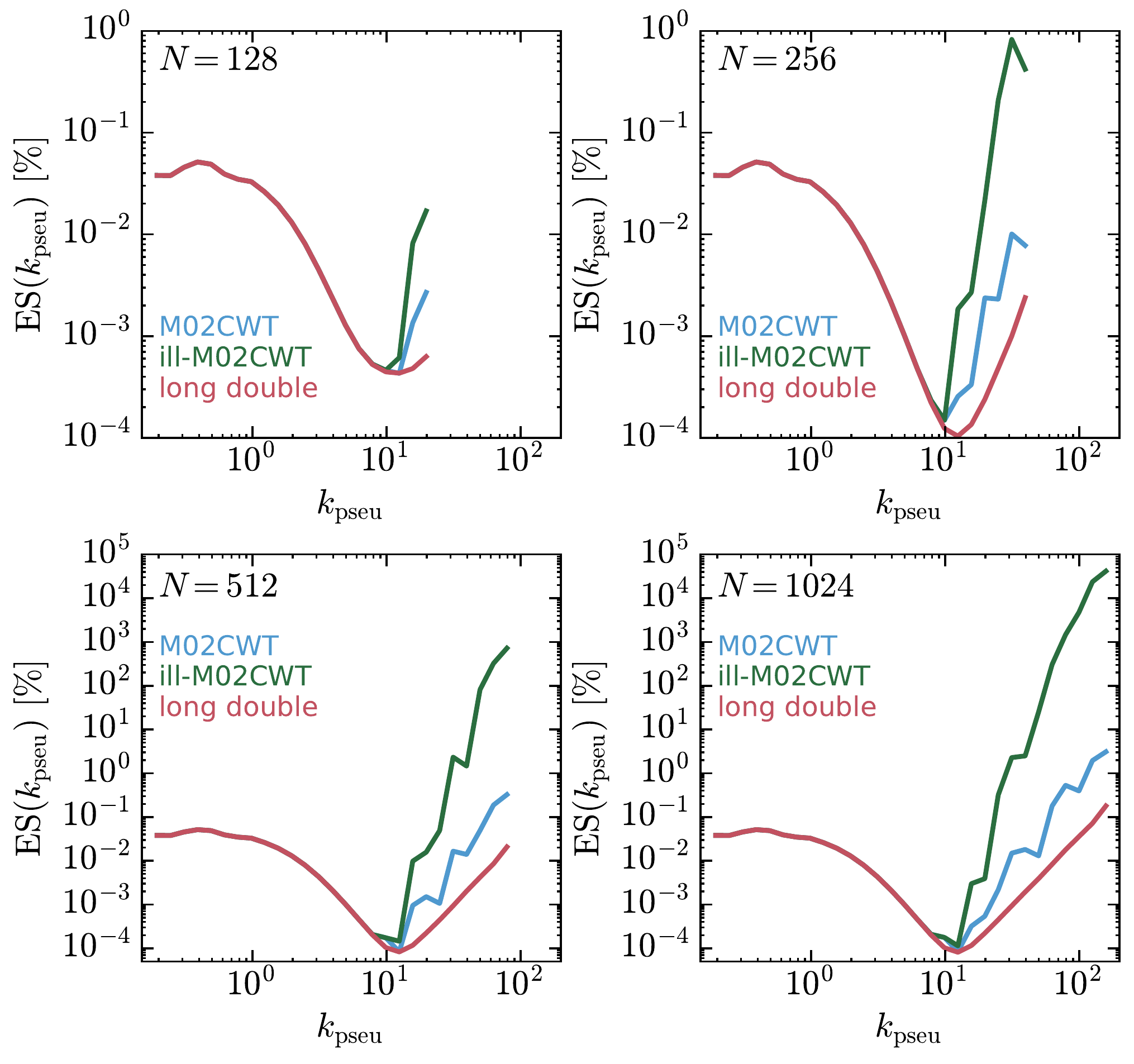}}
	\caption{The error spectra of the non-periodic signal $f_2(x)$, based on the CW-GDW at sampling numbers of $N=128$, $256$, $512$ and $1024$. The blue lines show the results obtained by the M02CWT algorithm with computing the sequence $c(l)$ locally at scale levels $i\geq 1$. The green lines, labeled as ``ill-M02CWT", show the results obtained by the variant of the M02CWT with computing the sequence $c(l)$ globally. The red lines, labeled as ``long double", show the results obtained by the long double implementation of the ill-M02CWT. }
	\label{fig:es_m02_illm02}
\end{figure}

\begin{figure}
	\centerline{\includegraphics[width=0.47\textwidth]{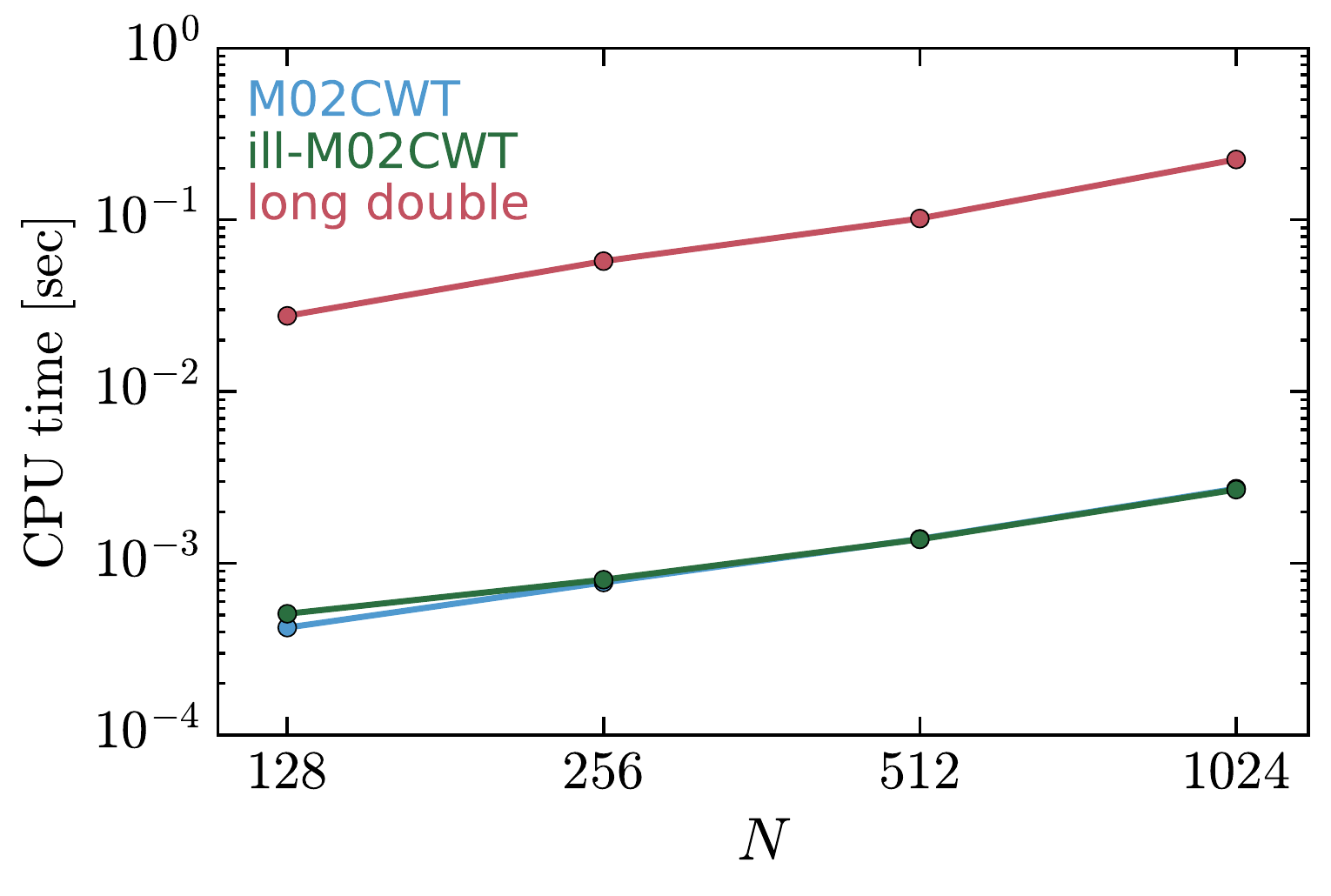}}
	\caption{The CPU time per scale of the M02CWT, ill-M02CWT, and the long double precision ill-M02CWT to compute the numerical CWT of the non-periodic signal $f_2(x)$ with the CW-GDW at sampling numbers of $N=128$, $256$, $512$ and $1024$.}
	\label{fig:cpu_time_m02_illm02}
\end{figure}

Both the M02CWT and A19CWT algorithms calculate cumulative sum of the coefficient sequence $c(l)$ four times (see Equation \eqref{eq:cumsum4_coeffs}). However, repeated cumulative summation can produce floating-points with huge values, which are less precise. Therefore, if we use the sequence of coefficients $g(l)$ that is computed in one go before the scale-dependent operations, i.e. Equations \eqref{eq:m02_cwt} and \eqref{eq:a19cwt_04}, then the algorithms will be terribly imprecise, which is not emphasized in \citet{Munoz2002} and \citet{Arizumi2019}.

As an example, we use the M02CWT algorithm to illustrate the accuracy issue, and denote its variant with computing $g(l)$ globally as the ill-M02CWT. In Fig. \ref{fig:es_m02_illm02}, we show that especially for the large sampling numbers of $N=512$ and $1024$, the ill-M02CWT yields very high errors at small scales. The M02CWT reduces the errors to a great extent. Although the errors of the long double precision ill-M02CWT are lower, the cost of using it is extremely expensive. As shown in Fig. \ref{fig:cpu_time_m02_illm02}, the CPU time consumed by the M02CWT is almost the same compared to the ill-M02CWT, while the long double precision ill-M02CWT takes tens of times more CPU time than the ill-M02CWT.

\bibliography{references}{}
\bibliographystyle{mnras}

\end{document}